\documentclass[10pt]{iopart}

    \usepackage{iopams}
    \usepackage{amssymb}
    \usepackage{amsfonts}
    \usepackage{amstext}
    \usepackage{graphicx}
    \usepackage{setstack}
    \usepackage{amscd}

    \newcommand{\ihbar}{\imath \hbar}
    \newcommand{\fig}[4]{\begin{figure}[h]
                         \begin{center}
                         \includegraphics*[width=#4]{#2}
                         \end{center}
                         \caption{\label{#1} \textit{#3}}
                         \end{figure}}
    \newcommand{\Te}{\mathbb{T}e}
    
    \newcommand{\mod}{\ \mathrm{mod}\ }

\begin{document}

    \title[Geometric phases in adiabatic Floquet theory]{Geometric phases in adiabatic Floquet theory, abelian gerbes and Cheon's anholonomy}

    \author{David Viennot\dag\footnote[7]{viennot@obs-besancon.fr}}

    \address{\dag\ Institut UTINAM (CNRS UMR 6213, Universit\'e de Franche-Comt\'e, Observatoire de Besan{\c c}on), 41 bis Avenue de
    l'Observatoire, BP1615, 25010 Besan{\c c}on cedex, France}

    \begin{abstract}
We study the geometric phase phenomenon in the context of the adiabatic Floquet theory (the so-called the $(t,t')$ Floquet theory). A double integration appears in the geometric phase formula because of the presence of two time variables within the theory. We show that the geometric phases are then identified with horizontal lifts of surfaces in an abelian gerbe with connection, rather than with horizontal lifts of curves in an abelian principal bundle. This higher degree in the geometric phase gauge theory is related to the appearance of changes in the Floquet blocks at the transitions between two local charts of the parameter manifold. We present the physical example of a kicked two-level system where these changes are involved a Cheon's anholonomy. In this context, the analogy between the usual geometric phase theory and the classical field theory also provides an analogy with the classical string theory.  
    \end{abstract}


\section{Introduction}
The Floquet theory introduced in quantum mechanics by Shirley  \cite{shirley}, is now a classical tool to treat time-periodic Hamiltonians. It is often used to describe quantum systems interacting with constant wave (cw) laser fields \cite{sambe,barone}. The adiabatic Floquet theory (so-called $(t,t')$ Floquet theory, which is a generalization of Shirley's works) and the related concept of quasi-energy are used to describe a quantum system interacting with a pulsed and chirped laser field \cite{guerin, drese}. They are also used to study kicked systems \cite{haake}, the control of quantum dynamics by laser fields \cite{guerin2} and other time-dependent phenomena \cite{chu}. The non-adiabatic geometric phases arising in the simple Floquet theory have been extensively studied by Moore and Stedman in \cite{moore1,moore2,moore3,moore4,moore5}. The non-adiabatic geometric phase phenomenon was discovered by Aharonov and Anandan in \cite{aharonov} as a sequel to the discovery of the adiabatic geometric phase phenomenon by Berry and Simon in \cite{berry,simon}. In the present work we study  both the non-adiabatic and the adiabatic geometric phases arising in the adiabatic Floquet theory. After a short overview of the Floquet theories, section 2 shows that the geometric phases involved by the adiabatic Floquet theory are generated by a double integration (rather than a simple integration in the usual geometric phase theory). Section 3 describes the geometric structure describing the geometric phases (an abelian gerbe with connection \cite{mackaay,picken,brylinski,hitchin}), and clarifies the signifiance of these double integrated geometric phases. We show in particular that this more complicated structure is related to  the existence of systems having a quasi-energy with non-global continuous definition of the Floquet blocks. Such a system is presented in section 4, which gives an illustrative example of the theoretical results of this paper.

\section{The Floquet theories and the associated geometric phases}
\subsection{The Floquet theory for a cw field or for a train of ultrashort pulses}
We consider a $\tau$-periodic time-dependent self-adjoint Hamiltonian $t \mapsto H(t)$ in the Hilbert space $\mathcal H$; for simplicity we consider that $\mathcal H$ is finite dimensional and so can be canonically identified with $\mathbb C^N$. We consider two interesting examples. The first one is the Hamiltonian corresponding to an atom or a molecule interacting with a cw laser field:
\begin{equation}
H^{ex1}(t) = H_0 + \mu E \cos(\omega t)
\end{equation}
where $H_0 \in \mathcal L(\mathcal H)$ is the free hamiltonian of the atom/molecule, $\mu \in \mathcal L(\mathcal H)$ is the dipolar moment of the atom/molecule, $E \in \mathbb R^+$ and $\omega = \frac{2 \pi}{\tau} \in \mathbb R^+$ are respectively the amplitude and the frequency of the laser field. The second example is the Hamiltonian of a kicked rotator, corresponding to an atom or a molecule interacting with a train of ultrashort pulses:
\begin{equation}
H^{ex2}(t) = H_0 + \hbar \lambda W \sum_{n \in \mathbb Z} \delta(t-n\tau) 
\end{equation}
where $H_0 \in \mathcal L(\mathcal H)$ is again the free hamiltonian of the atom/molecule, $W \in \mathcal L(\mathcal H)$ is the operator describing the effect of a kick on the atom/molecule, and $\lambda \in \mathbb R^+$ is the strength of a kick.\\
We introduce to the variable change $\theta = \omega t$, so that
\begin{equation}
H^{ex1}(\theta) = H_0 + \mu E \cos \theta
\end{equation}
or
\begin{equation}
H^{ex2}(\theta) = H_0 + \hbar \omega \lambda W \sum_{n \in \mathbb Z} \delta(\theta - 2 n \pi)
\end{equation}
with the Schr\"odinger equation
\begin{equation}
\ihbar \omega \frac{d \psi}{d\theta} = H(\theta) \psi(\theta)
\end{equation}

The Floquet theory can be expressed by using two equivalent formalisms. The first one, the Moore-Stedman formalism \cite{moore1,moore2,moore3,moore4}, considers the evolution operator $U(\theta) \in \mathcal U(\mathcal H)$ (where $\mathcal U(\mathcal H)$ is the set of unitary operators of $\mathcal H$). $U(\theta)$ obeys the equation
\begin{equation}
\ihbar \omega \frac{\partial U}{\partial \theta} = H(\theta) U(\theta) \qquad U(0) = id_{\mathcal H}
\end{equation}
By using the Floquet theorem, we can decompose the operator as follows:
\begin{equation}
U(\theta) = Z(\theta) e^{\imath \mathsf M \theta}
\end{equation}
where $Z(\theta) \in \mathcal U(\mathcal H)$ is a periodic unitary operator, with $Z(\theta + 2 \pi) = Z(\theta)$ and with $Z(0) = id_{\mathcal H}$, and where $\mathsf M \in \mathcal L(\mathcal H)$ is a constant self-adjoint operator. Let $\left\{ -\frac{\tilde \chi_j}{\hbar \omega} \right\}_{j=1,...,N}$ and $\left\{ |\mu_j \rangle  \in \mathcal H \right\}_{j=1,...,N}$ be, respectively, the eigenvalues (supposed non-degenerate) and the normalized eigenvectors of $\mathsf M$  
\begin{equation}
\mathsf M |\mu_j \rangle = -\frac{\tilde \chi_j}{\hbar \omega} |\mu_j \rangle 
\end{equation}
The Moore-Stedman Floquet formalism uses $(|\mu_j \rangle)_{j=1,...,N}$ as the basis of $\mathcal H$.\\

The second approach to the Floquet theory, the quasienergy formalism \cite{guerin2,chu} considers the Floquet hamiltonian
\begin{equation}
H_F = H(\theta) - \ihbar \omega \partial_\theta
\end{equation}
in the extended Hilbert space $\mathcal H \otimes \mathcal F$ where $\mathcal F = L^2\left( S^1, \frac{d\theta}{2 \pi} \right)$ is the space of square integrable functions on the circle $S^1$ ($H_F \in \mathcal L(\mathcal H \otimes \mathcal F)$ is self-adjoint). The extended Hilbert space is endowed with the scalar product
\begin{equation}
\forall \psi, \phi \in \mathcal H \otimes \mathcal F, \qquad \langle \psi | \phi \rangle_{\mathcal H \otimes \mathcal F} = \int_0^{2 \pi} \langle \psi(\theta)|\phi(\theta) \rangle_{\mathcal H} \frac{d\theta}{2 \pi}
\end{equation}
where $\langle \cdot | \cdot \rangle_{\mathcal H}$ is the scalar product on $\mathcal H$. Let $(|j \rangle)_{j =1,...,N}$ be an arbitrary basis of $\mathcal H$. Since $(e^{\imath n \theta})_{n \in \mathbb Z}$ is a basis of $\mathcal F$, we have
\begin{equation}
\forall \psi \in \mathcal H \otimes \mathcal F, \exists c_{j,n} \in \mathbb C, \qquad |\psi \rangle = \sum_{j = 1}^N \sum_{n \in \mathbb Z} c_{j,n} |j \rangle \otimes |e^{\imath n \theta} \rangle
\end{equation}
$\psi$ can be viewed as a $\theta$-dependent vector of $\mathcal H$ by writing
\begin{equation}
|\psi(\theta) \rangle = \sum_{j = 1}^N \left(\sum_{n \in \mathbb Z} c_{j,n} e^{\imath n \theta} \right) |j \rangle
\end{equation}
Let $\{\chi_a \}_{a \in \mathbb Z}$ and let $\left\{ |a \rangle \in \mathcal H \otimes \mathcal F \right\}_{a \in \mathbb Z}$ be, respectively, the eigenvalues and the normalized eigenvectors of $H_F$, so that
\begin{equation}
H_F|a \rangle = \chi_a |a \rangle
\end{equation}
$\{\chi_a \}_{a \in \mathbb Z}$ are called the quasienergies of the system. The quasienergy formalism uses $(| a \rangle )_{a \in \mathbb Z}$ as the basis of $\mathcal H \otimes \mathcal F$. The spectrum of $H_F$ is $\hbar \omega$-periodic, and the quasienergy state associated with $\chi_a + n \hbar \omega$ ($n \in \mathbb Z$) is the state $e^{\imath n \theta}|a \rangle$. We can consider the $N$ quasienergies with values in $[0, \hbar \omega[$ as forming the number 0 Floquet block, the $N$ quasienergies with values in $[\hbar \omega, 2 \hbar \omega[$ as forming the number 1 Floquet block, etc. This decomposition is arbitrary, and another possibility would be to continuously link a quasienergy with an eigenvalue of $H_0 - \ihbar \omega \partial_\theta \in \mathcal L(\mathcal H \otimes \mathcal F)$. Such an eigenvalue has the form $\chi_a^0 = \lambda_i + n \hbar \omega$ where $n \in \mathbb Z$ and $\lambda_i$ is one of the $N$ eigenvalues of $H_0$. If $\chi_a$ linked to $\chi_a^0 = \lambda_i + n \hbar \omega$, we say that it belongs to the number $n$ Floquet block, which can be physically interpreted as being the set of the quasienergies associated with $n$ photons exchanged between the atom/molecule and the laser field (see \cite{guerin3}).

The two formulations of the Floquet theory are related by
\begin{equation}
\forall a \in \mathbb Z, \exists j \in\{1,...,N\}, \exists n \in \mathbb Z, \text{ such that } \chi_a = \tilde \chi_j + n \hbar \omega
\end{equation}
\begin{equation}
|a(\theta) \rangle = e^{\imath n \theta} Z(\theta) |\mu_j \rangle, \text{ if } \chi_a = \tilde \chi_j + n \hbar \omega
\end{equation}
Note that $|a \rangle$, which is normalized in $\mathcal H \otimes \mathcal F$, is also normalized in $\mathcal H$: $\forall \theta$, $\langle a(\theta) |a(\theta) \rangle_{\mathcal H} = \langle \mu_i |Z^\dagger(\theta) e^{- \imath n \theta} e^{\imath n \theta} Z(\theta) |\mu_i \rangle_{\mathcal H} =1$.\\ 

Let $\theta \mapsto \psi(\theta) \in \mathcal H$ be the wave function defined by $\psi(\theta) = U(\theta) |\mu_j \rangle$, or equivalently let $\psi \in \mathcal H$ be the solution of the equation $H_F \psi = 0$ such that $\psi(0) = |\mu_j \rangle$. We have
\begin{equation}
\psi(2 \pi) = \underbrace{Z(2 \pi)}_{id_{\mathcal H}} e^{\imath \mathsf M 2 \pi} |\mu_j \rangle = e^{-\imath 2 \pi \frac{\tilde \chi_j}{\hbar \omega}} |\mu_j \rangle
\end{equation}  
However,
\begin{eqnarray}
& & H_F Z(\theta) |\mu_j \rangle  =  \tilde \chi_j Z(\theta) |\mu_j \rangle \\
& \iff & (H(\theta)- \ihbar \omega \partial_\theta) Z(\theta) |\mu_j \rangle  =  \tilde \chi_j Z(\theta) |\mu_j \rangle
\end{eqnarray}
Then, by projecting this last equation on $\langle \mu_j|Z^\dagger(\theta)$, we have
\begin{eqnarray}
\tilde \chi_j & = & \tilde \chi_j \int_0^{2 \pi} \underbrace{\langle \mu_j | Z(\theta)^\dagger Z(\theta) |\mu_j \rangle_{\mathcal H}}_{=1} \frac{d \theta}{2 \pi} \\
& = & \int_0^{2 \pi} \langle \mu_j |Z^\dagger(\theta) H(\theta) Z(\theta) |\mu_j \rangle_{\mathcal H} \frac{d\theta}{2 \pi} \nonumber \\
& & \qquad - \ihbar \omega \int_0^{2 \pi} \langle \mu_j |Z^\dagger(\theta) \frac{\partial Z(\theta)}{\partial \theta} |\mu_j \rangle_{\mathcal H} \frac{d\theta}{2 \pi}
\end{eqnarray}
Finally we have
\begin{eqnarray}
\psi(2 \pi) & = & e^{-\frac{\imath}{\hbar \omega}  \int_0^{2 \pi} \langle \mu_j |Z^\dagger(\theta) H(\theta) Z(\theta) |\mu_j \rangle_{\mathcal H} d\theta} \nonumber \\
& & \times e^{- \int_0^{2 \pi} \langle \mu_j |Z^\dagger(\theta) \frac{\partial Z(\theta)}{\partial \theta} |\mu_j \rangle_{\mathcal H} d \theta} |\mu_j \rangle
\end{eqnarray}
Moore and Stedman have pointed out \cite{moore1,moore2,moore3,moore4} that $e^{-\frac{\imath}{\hbar \omega}  \int_0^{2 \pi} \langle \mu_j |Z^\dagger(\theta) H(\theta) Z(\theta) |\mu_j \rangle_{\mathcal H} d\theta}$ constitutes a usual dynamical phase whereas $e^{- \int_0^{2 \pi} \langle \mu_j |Z^\dagger(\theta) \frac{\partial Z(\theta)}{\partial \theta} |\mu_j \rangle_{\mathcal H} d \theta}$ constitutes a geometric phase of a cyclic evolution, as defined by Aharonov and Anandan in \cite{aharonov}. 

\subsection{The adiabatic Floquet theory}
We consider now a parameter-dependent and time-dependent self-adjoint Hamiltonian $(\vec R,t) \mapsto H(\vec R,t) \in \mathcal L(\mathcal H)$. $H(\vec R,t)$ is supposed, moreover, to be $\frac{2\pi}{\omega}$-periodic in time where $\omega$ is (possibly) one of the parameters $\vec R$. We are interested in the dynamics generated by the parameter-modulated Hamiltonian $t \mapsto H(\vec R(t),t)$ where the modulation $t \mapsto \vec R(t)$ is slow with respect to the evolution rate associated with the explicit time-dependence of $H(\vec R,t)$. The two interesting examples become those cited in section 2.1, the Hamiltonian corresponding to an atom or a molecule interacting with a chirped laser field with envelope modulations
\begin{equation}
H^{ex1}(\vec R(t),t) = H_0 + \mu E(t) \cos(\omega(t)t)
\end{equation}
and the Hamiltonian corresponding to an atom or a molecule interacting with an irregular train of ultrashort pulses with different strengths:
\begin{equation}
H^{ex2}(\vec R(t),t) = H_0 + \hbar \lambda(t) W \sum_{n \in \mathbb Z} \delta\left(\omega(t)t- 2 n \pi \right)
\end{equation}
Let $\omega_0$ be a reference frequency and let $\phi(t) = \left(\omega(t)-\omega_0 \right)t \mod 2 \pi$ be the time-dependent phase of the frequency modulation. For convenience we use $\phi$ rather than $\omega$ as an adiabatic parameter within $\vec R$. We have
\begin{equation}
H^{ex1}(\vec R(t),t) = H_0 + \mu E(t) \cos\left(\omega_0 t + \phi(t) \right)
\end{equation}
with $\vec R=(E,\phi)$, and
\begin{equation}
H^{ex2}(\vec R(t),t) = H_0 + \hbar \lambda(t) W \sum_{n \in \mathbb Z} \delta\left(\omega_0t - 2 n \pi + \phi(t) \right)
\end{equation}
with $\vec R=(\lambda,\phi)$.\\

In order to separate the fast periodic terms from the slow adiabatic evolution generated by $t \mapsto \vec R(t)$ we introduce the new variable $\theta = \omega_0t$ and we consider the parameter-dependent Floquet Hamiltonian $\vec R \mapsto H_F(\vec R) \in \mathcal L(\mathcal H \otimes \mathcal F)$, defined for our two examples as
\begin{equation}
H_F^{ex1}(\vec R(t)) = H_0 + \mu E(t) \cos \left(\theta + \phi(t) \right) - \ihbar \omega_0 \partial_\theta
\end{equation}
and
\begin{equation}
H_F^{ex2}(\vec R(t)) = H_0 + \hbar \omega_0 \lambda(t) W \sum_{n \in \mathbb Z} \delta\left(\theta - 2 n \pi + \phi(t)\right) - \ihbar \omega_0 \partial_\theta
\end{equation}
By doing this we introduce a theory with two-time variables \cite{breuer,peskin}, both with a Floquet approach \cite{guerin,drese}.\\
Let $M$ be the $\mathcal C^\infty$-manifold generated by all configurations of the parameters $\vec R$. Let $\{U^\alpha \}_\alpha$ be a good open cover of $M$ (i.e. a set of contractible open sets of $M$ such that $\bigcup_\alpha U^\alpha = M$). Let $\{\chi_a \}_{a \in \mathbb Z}$ and $\{|a,\vec R \rangle^\alpha \in \mathcal H \otimes \mathcal F \}_{a \in \mathbb Z}$ be, respectively, the quasienergies and the quasienergy states on $U^\alpha$ of the $\vec R$-dependent Floquet Hamiltonian $H_F(\vec R)$. 
\begin{equation}
\forall \vec R \in U^\alpha, \quad H_F(\vec R)|a,\vec R \rangle^\alpha = \chi_a(\vec R) |a,\vec R \rangle^\alpha
\end{equation}
$\vec R \mapsto \chi_a(\vec R)$ is for the moment supposed continuous on the whole of $M$ and $\vec R \mapsto |a,\vec R \rangle^\alpha$ is supposed $\mathcal C^2$ on $U^\alpha$. The quasienergy states are locally defined (with one definition for each chart $U^\alpha$), because in general it is impossible to define a globally $\mathcal C^2$ eigenvector or to keep the same phase convention on the whole of $M$. Since $t \mapsto \vec R(t)$ represents a slow variation we can apply an adiabatic approximation \cite{messiah} to describe the solution of the Schr\"odinger equation
\begin{equation}
\left\{ \begin{array}{l} \ihbar \frac{\partial \psi}{\partial t} = H_F(\vec R(t)) \psi(t), \qquad \psi \in \mathcal H \otimes \mathcal F \\ \psi(0) = |a,\vec R(0) \rangle^\alpha \end{array} \right.
\end{equation}

Let $\mathcal C$ be the path in $M$ parametrized by $[0,T] \ni t \mapsto \vec R(t) \in M$. We suppose that $\chi_a$ is not degenerate on the whole of $M$ or at least that $\mathcal C$ does not pass in the proximity of the points of $M$ where $\chi_a$ crosses other quasienergies. If $\mathcal C \subset U^\alpha$ then we have
\begin{equation}
\psi(T) = e^{- \ihbar^{-1} \int_0^T \chi_a(\vec R(t')) dt'} e^{- \int_{\mathcal C} A^\alpha} |a,\vec R(T) \rangle^\alpha
\end{equation}
where
\begin{equation}
A^\alpha = {^\alpha}\langle a, \vec R|d_M|a,\vec R \rangle^\alpha_{\mathcal H \otimes \mathcal F} \in \Omega^1 U^\alpha
\end{equation}
$d_M$ being the exterior differential of $M$ and $\Omega^n U^\alpha$ being the set of differential $n$-forms of $U^\alpha$. $ e^{- \imath \gamma_a(\mathcal C)} = e^{- \int_{\mathcal C} A^\alpha}$ is the geometric phase of the adiabatic evolution as studied by Berry and Simon \cite{berry,simon}.\\
If now $\mathcal C$ passes through several charts, we have
\begin{equation}
\psi(T) = e^{- \ihbar^{-1} \int_0^T \chi_a(\vec R(t')) dt'} e^{- \imath \gamma_a(\mathcal C)} |a, \vec R(T) \rangle^\zeta
\end{equation}
where the geometric phase is defined by
\begin{equation}
\label{geophasemultichart}
e^{\imath \gamma_a(\mathcal C)} = e^{\int_{\vec R(0)}^{\vec R^{\alpha \beta}} A^\alpha} e^{\imath \varphi^{\alpha \beta}(\vec R^{\alpha \beta})} 
e^{\int_{\vec R^{\alpha \beta}}^{\vec R^{\beta \gamma}} A^\beta} e^{\imath \varphi^{\beta \gamma}(\vec R^{\beta \gamma})} ...
 e^{\int_{\vec R^{\xi \zeta}}^{\vec R(T)} A^\zeta}
\end{equation}
Here $\vec R^{\alpha \beta}$ is an arbitrary point in $U^\alpha \cap U^\beta \cap \mathcal C$, the integrations being along the path $\mathcal C$. The transition functions $e^{\imath \varphi^{\alpha \beta}}$ are defined by
\begin{equation}
\forall \vec R \in U^\alpha \cap U^\beta, \quad |a,\vec R \rangle^\beta = e^{\imath \varphi^{\alpha \beta}(\vec R)} |a,\vec R \rangle^\alpha
\end{equation}
Since $\forall \vec R \in U^\alpha \cap U^\beta$, $|a,\vec R\rangle^\alpha$ and $|a,\vec R \rangle^\beta$ are two normalized eigenvectors associated with the same non-degenerate eigenvalue $\chi_a(\vec R)$, they differ only by a phase factor $e^{\imath \varphi^{\alpha \beta}(\vec R)}$. The formula (\ref{geophasemultichart}) correctly defines the geometric phase, since the result is independent of the choice of arbitrary transition points $\{\vec R^{\alpha \beta} \}_{\alpha,\beta}$ as was proved by Alvarez \cite{alvarez} for a general abelian gauge theory. Since the transition functions satisfy the cocycle relations:
\begin{equation}
\label{cocycle_g}
\forall \vec R \in U^\alpha \cap U^\beta \cap U^\gamma, \quad  e^{\imath \varphi^{\alpha \beta}(\vec R)} e^{\imath \varphi^{\beta \gamma}(\vec R)} e^{\imath \varphi^{\gamma \alpha}(\vec R)} = 1
\end{equation}
\begin{equation}
\forall \vec R \in U^\alpha \cap U^\beta, \quad e^{\imath \varphi^{\beta \alpha}(\vec R)} = e^{- \imath \varphi^{\alpha \beta}(\vec R)}
\end{equation}
they define a principal $U(1)$-bundle ($U(1)$ denoting the group of complex numbers with unit modulus) endowed with a connection associated with the potential $A^\alpha$ (see \cite{nakahara} for a presentation of the principal bundle theory). The geometric phase is associated with the horizontal lift of $\mathcal C$ in this principal bundle (if $\mathcal C$ is closed, i.e. $\vec R(T) = \vec R(0)$, the geometric phase is the holonomy of the horizontal lift).\\

The parameter $\theta$ describing the fast evolution does not explicitly appear in the description of the geometric phase of the adiabatic Floquet theory, whereas it is the fundamental parameter in the description of the non-adiabatic geometric phase of the usual Floquet theory. We rewrite the previous expressions by viewing the states as $\theta$-dependent functions. First we have
\begin{eqnarray}
\chi_a(\vec R) & = & {^\alpha}\langle a,\vec R|H_F|a ,\vec R \rangle^\alpha_{\mathcal H \otimes \mathcal F} \\
& = & \int_0^{2 \pi} {^\alpha}\langle a(\theta),\vec R|H(\vec R,\theta)|a(\theta),\vec R \rangle^\alpha_{\mathcal H} \frac{d\theta}{2 \pi} \nonumber \\
& & \quad - \ihbar \omega_0  \int_0^{2 \pi} {^\alpha}\langle a(\theta),\vec R|\partial_\theta|a(\theta),\vec R \rangle^\alpha_{\mathcal H} \frac{d\theta}{2 \pi}
\end{eqnarray}
If $\mathcal C \subset U^\alpha$ we then have
\begin{eqnarray}
\psi(T) & = & e^{- \ihbar^{-1} \int_0^T \int_0^{2 \pi} {^\alpha} \langle a(\theta),\vec R(t')|H(\vec R(t'),t')|a(\theta),\vec R(t')\rangle^\alpha_{\mathcal H} \frac{d\theta}{2 \pi} dt'} \nonumber \\
& & \quad \times e^{- \int_0^T \int_0^{2 \pi} \eta_0^\alpha(\vec R(t'),\theta) d\theta dt'} e^{- \int_{\mathcal C} \int_0^{2 \pi} \eta_M^\alpha(\vec R,\theta) d\theta} |a,\vec R(T) \rangle^\alpha
\end{eqnarray}
where
\begin{equation}
\eta^\alpha_0 = \frac{\omega_0}{2 \pi} {^\alpha}\langle a(\theta),\vec R|\partial_\theta|a(\theta),\vec R \rangle^\alpha_{\mathcal H}
\end{equation}
\begin{equation}
\eta^\alpha_M = \frac{1}{2 \pi} {^\alpha}\langle a(\theta),\vec R|d_M|a(\theta),\vec R \rangle^\alpha_{\mathcal H}
\end{equation}
 $e^{- \int_0^T \int_0^{2 \pi} \eta_0^\alpha(\vec R(t'),\theta) d\theta dt'}$ is the geometric phase associated with the non-adiabatic fast cyclic evolution, whereas $e^{- \int_{\mathcal C} \int_0^{2 \pi} \eta_M^\alpha(\vec R,\theta) d\theta}$ is the geometric phase associated with the adiabatic slow evolution. We remark that these geometric phases are computed by a double integration. It then seems that they do not correspond to the horizontal lift of a curve. Moreover the geometric description must be constructed over the extended parameter manifold $M_+ = M \times S^1$, where $S^1$ is the circle parametrized by $\theta \mod 2\pi$ (since $|a(\theta+2\pi),\vec R \rangle^\alpha = |a(\theta),\vec R \rangle^\alpha$, the relevant extra dimension associated with $\theta$ is closed).\\
Another problem with the description presented in the begining of this section is that we have not taken into account the possibility that the Floquet block changes at the passage from one chart to another one. Indeed, we can imagine that the quasienergies are only locally defined, $\{\chi^\alpha_a \}_{a \in \mathbb Z}$ with
\begin{equation}
\forall \vec R \in U^\alpha, \quad H_F(\vec R) |a,\vec R \rangle^\alpha = \chi_a^\alpha(\vec R) |a,\vec R \rangle^\alpha
\end{equation}
$\vec R \mapsto \chi_a^\alpha(\vec R)$ being a continuous function on $U^\alpha$. At the passage from one chart to another one we have
\begin{equation}
\forall \vec R \in U^\alpha \cap U^\beta, \quad \chi_a^\beta(\vec R) = \chi_a^\alpha(\vec R) + n^{\alpha \beta} \hbar \omega_0 \text{ with } n^{\alpha \beta} \in \mathbb Z
\end{equation}
\begin{equation}
\forall \vec R \in U^\alpha \cap U^\beta, \quad |a(\theta),\vec R \rangle^\beta = e^{\imath \varphi^{\alpha \beta}(\vec R)} e^{\imath n^{\alpha \beta} \theta} |a(\theta),\vec R \rangle^\alpha
\end{equation}
We note that $n^{\alpha \beta}$ satisfies cocycle relations (we say that $n^{\alpha \beta}$ is $\delta$-closed):
\begin{equation}
n^{\alpha \beta} + n^{\beta \gamma} + n^{\gamma \alpha} = 0 \qquad \text{ if } U^\alpha \cap U^\beta \cap U^\gamma \not= \varnothing
\end{equation}
\begin{equation}
n^{\beta \alpha} = - n^{\alpha \beta} \qquad \text{ if } U^\alpha \cap U^\beta \not= \varnothing
\end{equation}
A trivial example producing a Floquet block transition arises when we compute the quasienergies by using the Moore-Stedman formalism with a local attribution of the Floquet blocks:
\begin{equation}
\chi^\alpha_a = \tilde \chi_i + p^\alpha \hbar \omega_0, \qquad p^\alpha \in \mathbb Z
\end{equation}
In this case, $n^{\alpha \beta} = p^\beta - p^\alpha$, and it is possible to cancel $n^{\alpha \beta}$ $\forall \alpha,\beta$ by redefining the Floquet block of each chart. However some systems have a particular topology such that $\forall p^\alpha \in \mathbb Z$ we have $n^{\alpha \beta} \not= p^\beta - p^\alpha$. For these systems it is impossible to redefining the Floquet blocks in order to cancel $n^{\alpha \beta}$. Section 4 presents such a system.\\
Taking into account the Floquet block changes associated with $n^{\alpha \beta}$, the wave function for a path $\mathcal C$ crossing several charts and finishing on the chart $U^\zeta$ is
\begin{equation}
\psi(T) =  e^{\imath \delta_a(T)} e^{- \imath \gamma_a(\mathcal S)} |a,\vec R(T) \rangle^\zeta
\end{equation}
where the dynamical phase is
\begin{equation}
e^{\imath \delta_a(T)} = e^{- \ihbar^{-1} \int_0^T \int_0^{2 \pi} {^\alpha} \langle a(\theta),\vec R(t')|H(\vec R(t'),t')|a(\theta),\vec R(t')\rangle^\alpha_{\mathcal H} \frac{d\theta}{2 \pi} dt'}
\end{equation}
and where the geometric phase is
\begin{eqnarray}
\label{phase_flo_adiab}
e^{\imath \gamma_a(\mathcal S)} & = & e^{\int_{\vec R(0)}^{\vec R^{\alpha \beta}} \int_0^{2 \pi} \eta^\alpha_M d\theta} e^{\int_0^{t^{\alpha \beta}} \int_0^{2 \pi} \eta^\alpha_0 d\theta dt} e^{\imath \varphi^{\alpha \beta}(\vec R^{\alpha \beta})} e^{\imath n^{\alpha \beta} \omega_0 t^{\alpha \beta}} ...\nonumber \\
& & ... e^{\imath \varphi^{\xi \zeta}(\vec R^{\xi \zeta})} e^{\imath n^{\xi \zeta} \omega_0 t^{\xi \zeta}} e^{\int_{\vec R^{\xi \zeta}}^{\vec R(T)} \int_0^{2 \pi} \eta^\zeta_M d\theta} e^{\int_{t^{\xi \zeta}}^{T} \int_0^{2 \pi} \eta^\zeta_0 d\theta dt}
\end{eqnarray}
$\mathcal S = \mathcal C \times S^1$ and $\vec R^{\alpha \beta} = \vec R(t^{\alpha \beta})$ is an arbitrary point in $\mathcal C \cap U^\alpha \cap U^\gamma$. The geometric integrations are along the path $\mathcal C$. Appendix A proves this formula.\\

The following section presents the geometric description of the geometric phases of the adiabatic Floquet theory, which involves the double integration, the extended parameter manifold and the changes of Floquet block at the chart transitions.

\section{The gerbe describing the geometric phases in the adiabatic Floquet theory}
Let $M_+^+ = M \times S^1 \times \mathbb R$ be the manifold of space-time parameters, where $\mathbb R$ models the set of times $t$. Let $\{U^\alpha\}_\alpha$ be a good open cover of $M$, $\{V^i\}_i$ be a good open cover of $S^1$ and $\{W^v\}_v$ be a good open cover of $\mathbb R$. $\{U^{[\alpha,i,v]} = U^\alpha \times V^i \times W^v \}_{\alpha,i,v}$ is then a good open cover of $M^+_+$ which should be used in the description. Nevertheless no relevant quantity depends on the indices $i$ and $v$; in order to simplify the notation and to clarify the discussion we omit these indices and adapt the formulae to use explicitly only $\{U^\alpha\}_\alpha$. 

\subsection{Connective structure and horizontal lift}
 We introduce the 2-form $B^\alpha \in \Omega^2(U^\alpha \times S^1 \times \mathbb R)$ defined by
\begin{eqnarray}
B^\alpha & = & \eta^\alpha_{M\mu}(\theta,\vec R) dR^\mu \wedge d\theta +F^\alpha_{M\mu \nu}(\theta,\vec R) dR^\mu \wedge dR^\nu \nonumber \\
& & \quad - \eta^\alpha_0(\theta,\vec R) d\theta \wedge dt - \frac{\partial \eta^\alpha_0(\theta,\vec R)}{\partial R^\mu} dR^\mu \wedge dt \nonumber \\
& & \quad + \frac{2 \pi}{\hbar \omega_0} \eta^\alpha_{M \mu}(\theta,\vec R) \frac{\partial \chi^\alpha_a(\vec R)}{\partial R^\nu} dR^\mu \wedge dR^\nu \nonumber \\
& & \quad - \frac{2 \pi}{\hbar \omega_0} \eta^\alpha_0(\theta,\vec R) \frac{\partial \chi^\alpha_a(\vec R)}{\partial R^\mu} dR^\mu \wedge dt
\end{eqnarray}
where the Einstein convention is adopted for the indices $\mu$ and $\nu$ from 1 up to the number of adiabatic parameters (the dimension of $M$), and where
\begin{equation}
F^\alpha_M = d_M \eta^\alpha_M \iff F^{\alpha}_{M \mu \nu} = \frac{1}{2} \left( \frac{\partial \eta^\alpha_{M\nu}}{\partial R^\mu}  - \frac{\partial \eta^\alpha_{M\mu}}{\partial R^\nu}  \right)
\end{equation}
We introduce also the 1-form $A^{\alpha \beta} \in \Omega^1(U^\alpha \cap U^\beta \times S^1 \times \mathbb R)$ such that
\begin{equation}
A^{\alpha \beta} = \frac{\imath}{2 \pi} \left(\varphi^{\alpha \beta}(\vec R)+n^{\alpha \beta} \omega_0t \right) \left(d\theta + \frac{2 \pi}{\hbar \omega_0} \frac{\partial \chi^\alpha_a(\vec R)}{\partial R^\mu} dR^\mu \right)
\end{equation}
Finally, we introduce the 0-form $h^{\alpha \beta \gamma} \in \Omega^0(U^\alpha \cap U^\beta \cap U^\gamma \times S^1 \times \mathbb R)$ defined by
\begin{equation}
\label{et1}
h^{\alpha \beta \gamma} = e^{- \imath \frac{2 \pi}{\hbar \omega_0} \chi^\alpha_a(\vec R) z^{\alpha \beta \gamma}} e^{- \imath z^{\alpha \beta \gamma} \theta}
\end{equation}
where $z^{\alpha \beta \gamma} \in \mathbb Z$ is defined by
\begin{equation}
\label{et2}
\forall \vec R \in U^\alpha \cap U^\beta \cap U^\gamma, \varphi^{\alpha \beta}(\vec R) + \varphi^{\beta \gamma}(\vec R)+ \varphi^{\gamma \alpha}(\vec R) = 2 \pi z^{\alpha \beta \gamma}
\end{equation}
This last equation arising from the complex logarithm of the equation (\ref{cocycle_g}).\\

By construction, we have the following equations (appendix B presents the details)
\begin{eqnarray}
\forall \vec R \in U^\alpha \cap U^\beta & \quad & d_{M^+_+} A^{\alpha \beta} = B^\beta - B^\alpha\\
\forall \vec R \in U^\alpha \cap U^\beta \cap U^\gamma & \quad & A^{\beta \gamma} - A^{\alpha \gamma} + A^{\alpha \beta} = - (h^{\alpha \beta \gamma})^{-1} d_{M^+_+}h^{\alpha \beta \gamma}\\
\forall \vec R \in U^\alpha & \quad & d_{M_+^+} B^\alpha = H
\end{eqnarray}
Here $H \in \Omega^3M_+^+$ is globally defined. These relations define an abelian gerbe endowed with a connective structure \cite{mackaay,picken,brylinski,hitchin}.

The abelian gerbe structure is the higher order generalization of the abelian principal bundle structure. Indeed the horizontal lift of a curve is naturally defined within a principal bundle, whereas the horizontal lift of a surface is naturally defined within a gerbe \cite{mackaay}. The general formula for ``the phase'' of the horizontal lift of a surface $\mathcal S$ (the holonomy of the horizontal lift if $\mathcal S$ is closed) was originally proposed by Alvarez in \cite{alvarez}. In this paper we focus on the particular case of the situation exposed in the previous section for a dynamics associated with a path $\mathcal C$ in $M$. Let $f$ be the immersion map associated with the dynamics:
\begin{equation}
f : \begin{array}{rcl} \mathbb [0,T] \times S^1 & \to & M \times S^1 \times \mathbb R \\ (t,\theta) & \mapsto & (\vec R(t),\theta,t) \end{array}
\end{equation}
Considering the horizontal lift of the surface $\mathcal S = f([0,T] \times S^1)$ in the gerbe, we have
\begin{equation}
\label{formule_gnle_phase}
e^{\imath \gamma_a(\mathcal S)} = \prod_\alpha e^{\int \hspace{-0.5em} \int_{\sigma^\alpha} f^* B^\alpha} \prod_{\alpha \beta} e^{\int_{E^{\alpha \beta}} f^* A^{\alpha \beta}} 
\end{equation}
$f^* :\Omega^n (M_+^+)  \to  \Omega^n([0,T] \times S^1) $ is the pull-back map of $f$. $\sigma^\alpha$ is a surface on $[0,T] \times S^1$ (with the same orientation) such that $\sigma^\alpha \subset f^{-1}(U^\alpha \times S^1 \times [0,T])$ and such that $\bigcup_\alpha \sigma^\alpha = [0,T] \times S^1$. Also $E^{\alpha \beta} = f^{-1}(\{\vec R^{\alpha \beta} \} \times S^1 \times \{t^{\alpha \beta}\})$ where $\vec R^{\alpha \beta}$ is an arbitrary point on $\mathcal C \cap U^\alpha \cap U^\beta$ and $t^{\alpha \beta}$ is such that $\vec R(t^{\alpha \beta}) = \vec R^{\alpha \beta}$. The products are such that $\alpha$ follows the indices of charts crossed by $\mathcal S$. Figure \ref{sheet} presents the geometric situation.
\fig{sheet}{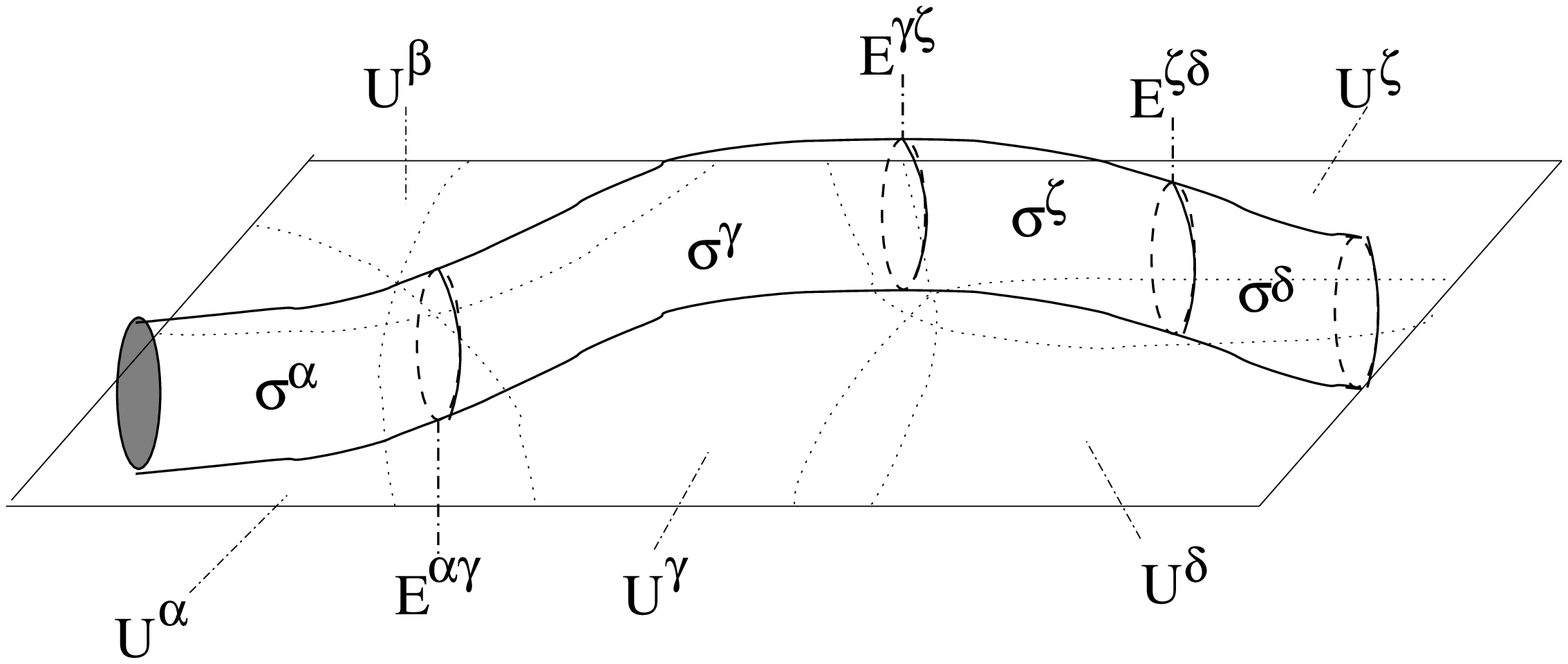}{Scheme of the sheet associated with the dynamics $\mathcal C \times S^1$. The plane represents the parameter manifold $M$ endowed with its chart system $\{U^\alpha \}_\alpha$. The sheet homeomorphic to a cylinder represents $\mathcal S$, the image of $[0,T] \times S^1$ in $M_+ = M \times S^1$, with its partition $\{\sigma^\alpha\}_\alpha$ and its transition paths $\{E^{\alpha \beta}\}_{\alpha,\beta}$.}{10cm}

We compute the pull-backs of $B^\alpha$ and of $A^{\alpha \beta}$ :
\begin{equation}
f^* B^\alpha = \eta^\alpha_{M\mu}(\theta,\vec R(t)) \frac{\partial \vec R^\mu}{\partial t} dt \wedge d\theta - \eta^\alpha_0(\theta,\vec R(t)) d\theta \wedge dt
\end{equation}
\begin{equation}
f^* A^{\alpha \beta} =  \frac{\imath}{2 \pi} \left(\varphi^{\alpha \beta}(\vec R(t)) + n^{\alpha \beta} \omega_0 t \right) \left(d\theta+ \frac{2 \pi}{\hbar \omega_0} \frac{d\chi^\alpha_a(\vec R(t))}{dt} dt \right)
\end{equation}

Since $\forall \alpha$, $\exists t^{\xi \alpha},t^{\alpha \zeta} \in [0,T]$ such that $\sigma^\alpha = [t^{\xi \alpha},t^{\alpha \zeta}] \times S^1$, we have
\begin{eqnarray}
& & \int \hspace{-0.5em} \int_{\sigma^\alpha} f^* B^\alpha \nonumber \\
& & \quad  =  \int_{t^{\xi \alpha}}^{t^{\alpha \zeta}} \int_0^{2 \pi} (\eta^\alpha_{M\mu}(\theta,\vec R(t')) \frac{\partial \vec R^\mu}{\partial t'} + \eta^\alpha_0(\theta,\vec R(t))) dt' \wedge d\theta \\
& & \quad  =  \int_{\vec R^{\xi \alpha}}^{\vec R^{\alpha \zeta}} \int_0^{2 \pi} \eta^\alpha_M(\vec R,\theta) d\theta + \int_{t^{\xi \alpha}}^{t^{\alpha \zeta}} \int_0^{2\pi} \eta^\alpha_0(\theta,\vec R(t'))) dt' d\theta
\end{eqnarray}
The integration from $\vec R^{\xi \alpha} = \vec R(t^{\xi \alpha})$ to $\vec R^{\alpha \zeta} = \vec R(t^{\alpha \zeta})$ is along $\mathcal C$. We have
\begin{eqnarray}
\int_{E^{\alpha \beta}} f^* A^{\alpha \beta} & = & \frac{\imath}{2 \pi} \int_0^{2\pi} \left(\varphi^{\alpha \beta}(\vec R^{\alpha \beta}) + \omega_0 t^{\alpha \beta} \right) d\theta \\
& =&  \imath \left(\varphi^{\alpha \beta}(\vec R^{\alpha \beta}) + \omega_0 t^{\alpha \beta} \right)
\end{eqnarray}
Finally we see that the equation (\ref{formule_gnle_phase}) coincides with the equation (\ref{phase_flo_adiab}). We conclude that the connective structure of the gerbe involves the geometric phase of the adiabatic Floquet theory. In accordance with the double integration, the geometric phase of the adiabatic Floquet theory is associated with the horizontal lift of a surface $\mathcal S = \mathcal C \times S^1$ rather than of a curve.\\

It is well known that there exists an analogy between the adiabatic geometric phase theory and the classical field theory (see for example \cite{bohm, wilczek}). The wave function with an adiabatic geometric phase $\psi(T) = e^{- \imath \oint_{\mathcal C} A} |a,\vec R(0) \rangle$ (omitting the dynamical phase and the question of chart transitions) is similar to the wave function of a charged particle within the space $M$, moving along the trajectory $\mathcal C$ and interacting with the magnetic field $F=d_MA$. Moreover $F$ is generated by magnetic monopoles within $M$ and associated with the crossings of $E_a$ with other eigenvalues. The theory of geometric phases in the adiabatic Floquet theory is similar to the classical string theory. The geometric phase is similar to $e^{\imath S}$, where $S$ is the world-sheet action of a charged closed string \cite{bonora,freed,kapustin} within the extended space-time $M_+^+ = M \times S^1 \times \mathbb R$. The string moves along the world-sheet $\mathcal S = f([0,T] \times S^1)$ and interacts with the Neveu-Schwarz B-field $B$ (a world-sheet is the two-dimensional generalization of a worldline). We note that the ``magnetic part'' of $B$ (components $dR^\mu \wedge dR^\nu$ and $dR^\mu \wedge d\theta$) is associated with the adiabatic evolution whereas the ``electric part'' of $B$ (components $dR^\mu \wedge dt$ and $d\theta \wedge dt$) is associated with the non-adiabatic fast cyclic evolution.

\subsection{Gauge transformations and topology of the gerbe}
Usually a gerbe connective structure obeys the following gauge transformations:
\begin{eqnarray}
\label{gauge_deg2}
\tilde h^{\alpha \beta \gamma} & = & h^{\alpha \beta \gamma} g^{\beta \gamma} (g^{\alpha \gamma})^{-1} g^{\alpha \beta} \\
\tilde A^{\alpha \beta} & = & A^{\alpha \beta} + (g^{\alpha \beta})^{-1} d_{M^+_+} g^{\alpha \beta} + k^\beta-k^\alpha \\
\tilde B^\alpha & = & B^\alpha + d_{M^+_+} k^\alpha
\end{eqnarray}
where $g^{\alpha \beta}$ is a $U(1)$-valued function and where $k^\alpha$ is a 1-form. Such transformations are still formally possible, but physically the gauge transformations must preserve the quasienergy states. The physically acceptable gauge transformations are then defined by
\begin{equation}
\widetilde{|a(\theta),\vec R \rangle^\alpha} = e^{\imath \epsilon^{\alpha}(\vec R)} e^{\imath p^\alpha \theta} |a(\theta),\vec R\rangle^\alpha
\end{equation}
where $\epsilon^\alpha$ is a function defined on $U^\alpha$ and $p^\alpha \in \mathbb Z$. This gauge transformation consists of a change in the arbitrary phase of the eigenvector of $H_F$ and in the Floquet block of the quasienergy. Moreover we can redefine $\varphi^{\alpha \beta}$ by adding $2 \pi m^{\alpha \beta}$ ($m^{\alpha \beta} \in \mathbb Z$) without modifying $e^{\imath \varphi^{\alpha \beta}}$. Under these two transformations we have
\begin{eqnarray}
\tilde \varphi^{\alpha \beta} & = & \varphi^{\alpha \beta} + \epsilon^\beta - \epsilon^\alpha + 2 \pi m^{\alpha \beta}\\
\tilde n^{\alpha \beta} & = & n^{\alpha \beta} + p^\beta - p^\alpha \\
\tilde \chi^\alpha_a & = & \chi^\alpha_a + p^\alpha \hbar \omega_0 \\
\tilde \eta^\alpha_M & = & \eta^\alpha_M + \frac{\imath}{2 \pi} d_M \epsilon^\alpha \\
\tilde \eta^\alpha_0 & = & \eta^\alpha_0 + \frac{\imath}{2 \pi} p^\alpha \omega_0 \\
\tilde B^\alpha & = & B^\alpha + \frac{\imath}{2 \pi} d_M \epsilon^\alpha \wedge \left(d\theta + \frac{2\pi}{\hbar \omega_0} d_M \chi^\alpha_a \right) \nonumber \\
 & & \quad - \frac{\imath}{2 \pi} p^\alpha \omega_0 \left(d\theta+ \frac{2 \pi}{\hbar \omega_0}d_M \chi^\alpha_a\right) \wedge dt \\
\tilde A^{\alpha \beta} & = & A^{\alpha \beta} \nonumber + \frac{\imath}{2 \pi} (\epsilon^\beta-\epsilon^\alpha +(p^\beta-p^\alpha) \omega_0t + 2 \pi m^{\alpha \beta}) \nonumber \\
& & \qquad \times \left(d\theta+ \frac{2 \pi}{\hbar \omega_0} d_M \chi^\alpha_a \right) \\
\tilde h^{\alpha \beta \gamma} & = & h^{\alpha \beta \gamma} e^{\imath (m^{\beta \gamma} - m^{\alpha \gamma} + m^{\alpha \beta}) (\theta+\frac{2 \pi}{\hbar \omega_0}\chi^\alpha_a)}
\end{eqnarray}
This corresponds to the gauge transformations (\ref{gauge_deg2}) but with a restriction on $g^{\alpha \beta}$ and on $k^\alpha$ which must be of the following form
\begin{equation}
g^{\alpha \beta} = e^{\imath m^{\alpha \beta} (\theta+\frac{2 \pi}{\hbar \omega_0} \chi^\alpha_a)} \in \Omega^0(U^\alpha \cap U^\beta \times S^1 \times \mathbb R) \qquad m^{\alpha \beta} \in \mathbb Z
\end{equation}
\begin{eqnarray}
k^{\alpha} & = & \frac{\imath}{2 \pi} (\epsilon^\alpha(\vec R) + p^\alpha \omega_0 t) \left(d \theta+\frac{2 \pi}{\hbar \omega_0} d_M \chi^\alpha_a\right) \in \Omega^1(U^\alpha \times S^1 \times \mathbb R) \nonumber \\
& & \hspace{7cm} p^\alpha \in \mathbb Z
\end{eqnarray}
To preserve the physical meaning it is necessary to restrict the gauge choices to these transformations.\\

The topology of a gerbe endowed with a connective structure is characterized by a Dixmier-Douady class $\check dd \in \check H^3(M^+_+, \mathbb Z)$, where $\check H^n(M^+_+, \mathbb Z)$ is $n$-th integer valued \v Cech cohomology group  (see \cite{brylinski}). A definition of the Dixmier-Douady class is the following. Let $w^{\alpha \beta \gamma \delta} \in \mathbb Z$ be such that
\begin{equation}
\label{et3}
\ln h^{\beta \gamma \delta} - \ln h^{\alpha \gamma \delta} + \ln h^{\alpha \beta \delta} - \ln h^{\alpha \beta \gamma} = -2 \pi \imath w^{\alpha \beta \gamma \delta}
\end{equation}
and let $[w]$ be the equivalence class of $w^{\alpha \beta \gamma}$ defined by
\begin{equation}
[w] = \left\{w^{\alpha \beta \gamma \delta} + x^{\beta \gamma \delta} - x^{\alpha \gamma \delta} + x^{\alpha \beta \delta} - x^{\alpha \beta \gamma}; x^{\alpha \beta \gamma} \in \mathbb Z \right\}
\end{equation}
At the inductive limit of the refinement of the good cover $\{U^{[\alpha,i,v]}\}_{\alpha,i,v}$, $[w]$ tends to $\check dd \in \check H^3(M^+_+, \mathbb Z)$ (see the books \cite{brylinski,bott} for a complete exposition of the \v Cech cohomology theory). In the present case
\begin{eqnarray}
-2 \pi \imath w^{\alpha \beta \gamma \delta} & = & - \imath \frac{2 \pi}{\hbar \omega_0}\left(\chi^\beta_a z^{\beta \gamma \delta} - \chi^\alpha_a (z^{\alpha \gamma \delta}-z^{\alpha \beta \delta}+z^{\alpha \beta \gamma}) \right) \nonumber \\
& & \quad - \imath \left(z^{\beta \gamma \delta}-z^{\alpha \gamma \delta}+z^{\alpha \beta \delta}-z^{\alpha \beta \gamma} \right) \theta
\end{eqnarray}
By using equation (\ref{et2}) we find that $z^{\beta \gamma \delta}-z^{\alpha \gamma \delta}+z^{\alpha \beta \delta}-z^{\alpha \beta \gamma} = 0$. We have then
\begin{eqnarray}
w^{\alpha \beta \gamma \delta} & = & \frac{1}{\hbar \omega_0}(\chi^\beta_a - \chi^\alpha_a) z^{\beta \gamma \delta} \\
& = & n^{\alpha \beta} z^{\beta \gamma \delta}
\end{eqnarray}
The Dixmier-Douady class is then the ``cup-product'' of two lower classes. Let $[z]$ be the equivalence class of $z^{\alpha \beta \gamma}$ defined by
\begin{equation}
[z] = \{z^{\alpha \beta \gamma} + x^{\beta \gamma} - x^{\alpha \gamma} + x^{\alpha \beta};x^{\alpha \beta} \in \mathbb Z \}
\end{equation}
At the inductive limit of the refinement of $\{U^\alpha\}_\alpha$, $[z]$ tends to $\check c_1 \in \check H^2(M,\mathbb Z)$. $\check c_1$ is the first Chern class of the principal $U(1)$-bundle defined by the transition functions $e^{\imath \varphi^{\alpha \beta}}$ (the bundle describes the pure adiabatic geometric phase $e^{- \int_{\mathcal C} {^\alpha}\langle a,\vec R|d_M|a,\vec R \rangle^\alpha_{\mathcal H \otimes \mathcal F}}$). It is well known that the first Chern class characterizes the non-trivial topology of the principal bundle.\\
Let $[n]$ be the equivalence class of $n^{\alpha \beta}$ defined by
\begin{equation}
[n] = \{n^{\alpha \beta} + p^\beta - p^\alpha; p^\alpha \in \mathbb Z \}
\end{equation}
At the inductive limit of the refinement of $\{U^\alpha\}_\alpha$, $[n]$ tends to $\check \nu \in \check H^1(M,\mathbb Z)$. $\check \nu$ characterizes the non-triviality of the quasienergy (and consequently the non-triviality of the non-adiabatic geometric phase phenomenon). The next section shows that this non-triviality is associated with the Cheon's anholonomy.\\
The two classes $\check \nu \in \check H^1(M,\mathbb Z)$ and $\check c_1 \in \check H^2(M,\mathbb Z)$ (or their cup-product $\check dd \in \check H^3(M,\mathbb Z)$) capture the topology of the gerbe associated with the adiabatic Floquet theory.\\
Remark: $\check dd$, $\check c_1$ and $\nu$ can be represented by group cohomology classes, $[h]=\{h^{\alpha \beta \gamma} g^{\beta \gamma} (g^{\alpha \gamma})^{-1} g^{\alpha \beta},g^{\alpha \beta}:U^\alpha \cap U^\beta \to U(1)\}$ is at the inductive limit of the refinement an element of $H^2(M,U(1))$, $[e^{\imath \varphi}] = \{e^{\imath \varphi^{\alpha \beta}} e^{\imath \epsilon^\beta} e^{- \imath \epsilon^\alpha}; \epsilon^\alpha:U^\alpha \to \mathbb R\}$ is at the inductive limit of the refinement an element of $H^1(M,U(1))$; and $[e^{\imath \chi_a}] = \{ e^{\imath \frac{2 \pi}{\hbar \omega_0} \chi_a^\alpha} e^{\imath \phi}, \phi : M \to \mathbb R \}$ is at the inductive limit of the refinement an element of $H^0(M,U(1))$.

\section{Example : a kicked two-level system exhibiting a Cheon's anholonomy}
The Cheon's anholonomy was originally discovered in the context of one-dimensional quantum systems submitted to pointlike potentials \cite{cheon}. Miyamoto and Tanaka showed \cite{miyamoto} the existence of Cheon's anholonomies in the context of the Floquet theory. To give a practical illustration of the theory of the present paper, we use the example treated in \cite{miyamoto}.

\subsection{The model}
We consider the Hamiltonian of a two-level system interacting with a regular train of ultrashort pulses:
\begin{equation}
\label{model}
H(\lambda,\theta) = H_0 + \hbar \omega_0 \lambda W \sum_{n \in \mathbb Z} \delta(\theta - 2 n \pi)
\end{equation}
where $\theta = \omega_0 t$ and
\begin{equation}
H_0 = \frac{\hbar \omega_1}{2} |\downarrow \rangle \langle \downarrow |
\end{equation}
within the Hilbert space $\mathcal H$ spanned by $\{ |\uparrow \rangle, |\downarrow \rangle \}$. The kick operator $W$ is the following rank one projector
\begin{equation}
W = |w \rangle \langle w | \qquad |w \rangle = \frac{1}{\sqrt 2} \left(|\uparrow \rangle - \imath | \downarrow \rangle \right)
\end{equation}
In order to simplify the discussion and to focus on the topology associated with the Floquet block changes, the kick strength is the only parameter which will be adiabatically modulated, i.e. $\vec R = \lambda$. The frequency of kicks $\omega_0$ will be kept constant. \\
Let $U_\lambda(\theta) \in \mathcal U(\mathcal H)$ be the solution of
\begin{equation}
\ihbar \omega_0 \frac{\partial U_\lambda}{\partial \theta} = H(\lambda,\theta) U_\lambda(\theta) \quad ; \quad  U_\lambda(0) = id_{\mathcal H}
\end{equation}
We can prove (see appendix C) that
\begin{equation}
U_\lambda(\theta) = \left\{ \begin{array}{ll} e^{- \frac{\imath}{\hbar \omega_0} H_0 \theta} & \text{if } \theta \in [0,2\pi[ \\
 \left(id_{\mathcal H} +\left(e^{- \imath \lambda}-1 \right) W \right) e^{- \frac{\imath}{\hbar \omega_0} H_0 2\pi} & \text{if } \theta = 2 \pi
\end{array} \right.
\end{equation}

We note that $\forall \theta$, $U_{\lambda + 2 \pi}(\theta) = U_\lambda(\theta)$. This implies that the quasienergy states $|a(\theta),\lambda \rangle = e^{\imath n \theta} Z_\lambda(\theta)|\mu_j,\lambda \rangle$ (where $|\mu_j,\lambda \rangle$ is an eigenvector of $\mathsf M_\lambda$, $\mathsf M_\lambda$ and $Z_\lambda(\theta)$ being the operators in the Floquet decomposition of $U_\lambda(\theta)$) are $2 \pi$-periodic with respect to $\lambda$. We conclude (for the moment) that the relevant parameter space is $M = S^1$, the circle parametrized by $\lambda \mod 2 \pi$.

\subsection{Quasienergies and Cheon's anholonomy}
We compute the quasienergies by using the Moore-Stedman formalism.
\begin{equation}
e^{\imath 2 \pi \mathsf M_\lambda} = U_\lambda(2 \pi)
\end{equation}
\begin{equation}
e^{- \frac{\imath}{\hbar \omega_0} H_0 2 \pi} = |\uparrow \rangle \langle \uparrow | + e^{- \imath \pi \frac{\omega_1}{\omega_0}} |\downarrow \rangle \langle \downarrow|
\end{equation}
We then have
\begin{equation}
U_\lambda(2 \pi) = \frac{1}{2} \left( \begin{array}{cc} e^{- \imath \lambda}+1 & \imath (e^{- \imath \lambda}-1)e^{-\imath \pi \frac{\omega_0}{\omega_1}} \\ - \imath (e^{-\imath \lambda}-1) & (e^{- \imath \lambda}+1)e^{-\imath \pi \frac{\omega_1}{\omega_0}} \end{array} \right)_{(|\uparrow \rangle,|\downarrow \rangle)}
\end{equation}
We first consider the particular case $\omega_0 = \omega_1$, so that
\begin{equation}
U_\lambda(2 \pi) = \frac{1}{2} \left( \begin{array}{cc} e^{- \imath \lambda} +1 & -\imath(e^{- \imath \lambda}-1) \\ -\imath(e^{- \imath \lambda}-1) & -(e^{-\imath \lambda}+1) \end{array} \right)_{(|\uparrow \rangle,|\downarrow \rangle)}
\end{equation}
We have
\begin{equation}
U_\lambda(2 \pi) |\mu_1,\lambda \rangle = e^{- \imath \frac{\lambda}{2}} |\mu_1,\lambda \rangle \quad , \quad |\mu_1,\lambda \rangle = \cos \frac{\lambda}{4} |\uparrow \rangle - \sin \frac{\lambda}{4} |\downarrow \rangle
\end{equation}
\begin{equation}
U_\lambda(2 \pi) |\mu_2,\lambda \rangle = -e^{- \imath \frac{\lambda}{2}} |\mu_2,\lambda \rangle \quad , \quad |\mu_2,\lambda \rangle = \sin \frac{\lambda}{4} |\uparrow \rangle + \cos \frac{\lambda}{4} |\downarrow \rangle
\end{equation}
We then have
\begin{equation}
- \frac{\tilde \chi_1(\lambda)}{\hbar \omega_0} 2\pi = - \frac{\lambda}{2} \qquad - \frac{\tilde \chi_2(\lambda)}{\hbar \omega_0} 2\pi = - \frac{\lambda}{2} - \pi
\end{equation}
Let $\{\chi_a \}_{a \in \mathbb Z}$ be the quasienergies defined by
\begin{eqnarray}
\chi_{2n+1}(\lambda) & = & \tilde \chi_1(\lambda) +n \hbar \omega_0 \qquad \forall n \in \mathbb Z\\
\chi_{2n+2}(\lambda) & = & \tilde \chi_2(\lambda)+n \hbar \omega_0 \qquad \forall n \in \mathbb Z
\end{eqnarray}
Then $\forall \lambda \in [0,2 \pi]$ we have
\begin{eqnarray}
\chi_1(\lambda) & = & \lambda \frac{\hbar \omega_0}{4 \pi} \\
\chi_2(\lambda) & = & \lambda \frac{\hbar \omega_0}{4 \pi} + \frac{\hbar \omega_0}{2}
\end{eqnarray}
We note a disconnection at $\lambda = 2 \pi$ if we follow by continuity the quasienergies with respect to $\lambda$:
\begin{eqnarray}
\chi_1(0) = 0 & \qquad & \chi_1(2 \pi) = \frac{\hbar \omega_0}{2} = \chi_2(0) \\
\chi_2(0) = \frac{\hbar \omega_0}{2} & \qquad & \chi_2(2 \pi) = \hbar \omega_0 = \chi_3(0) = \chi_1(0) + \hbar \omega_0
\end{eqnarray}
This effect associated with the exchange of $\tilde \chi_1$ and $\tilde \chi_2$ when using the continuity following of $\lambda \in [0,2\pi]$, is the Cheon's anholonomy illustrated in figure \ref{donut}.
\fig{donut}{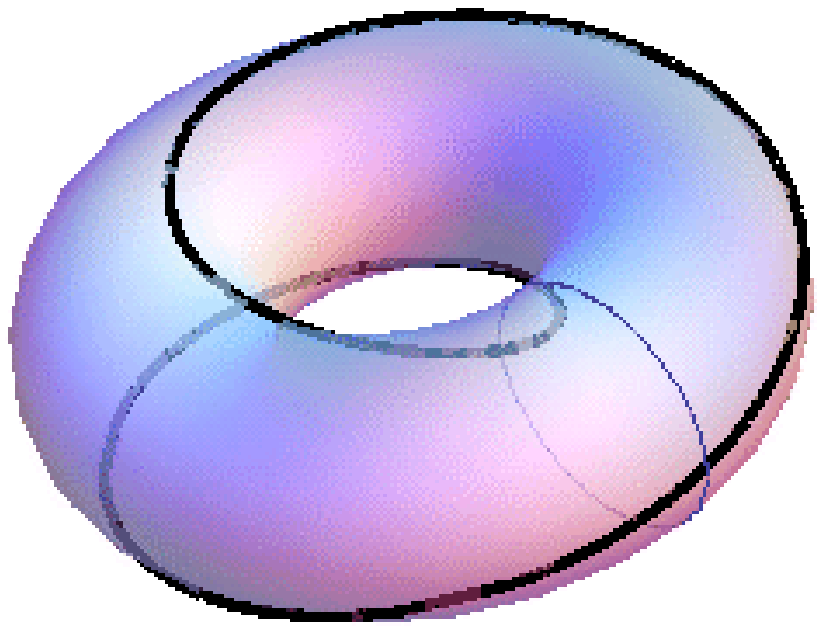}{Representation of the Cheon's anholonomy of the quasienergies of the Hamiltonian (\ref{model}) for $\omega_0 = \omega_1$. The grand circle of the torus is $\lambda \mod 2 \pi$ and the section little circles are the quasienergy space modulo $\hbar \omega_0$. The path represents the trajectories $\lambda \mapsto \tilde \chi_1(\lambda)$ and $\lambda \mapsto \tilde \chi_2(\lambda)$. The section circle corresponding to $\lambda = 0 \mod 2 \pi$ is drawn on the torus.}{6cm} 
In order to restore a sort of continuity for the quasienergies with respect to the adiabatic parameter, we use $[0,4\pi]$ rather than $[0,2\pi]$ as the range of $\lambda$:
\begin{eqnarray}
\chi_1(0) = 0 & \qquad & \chi_1(4 \pi) = \chi_1(0) + \hbar \omega_0 = \chi_3(0) \\
\chi_2(0) = \frac{\hbar \omega_0}{2} & \qquad & \chi_2(4 \pi) = \chi_2(0) + \hbar \omega_0 = \chi_4(0)
\end{eqnarray}
The quasienergies are then continuous (modulo a Floquet block change). In fact the system presents a Cheon's anholonomy for all values of $\omega_0$, except for $\omega_0 = \frac{\omega_1}{2}$, where a quasienergy crossing occurs at $\lambda = 0 \mod 2 \pi$ (see figure \ref{quasienergpath}).
\fig{quasienergpath}{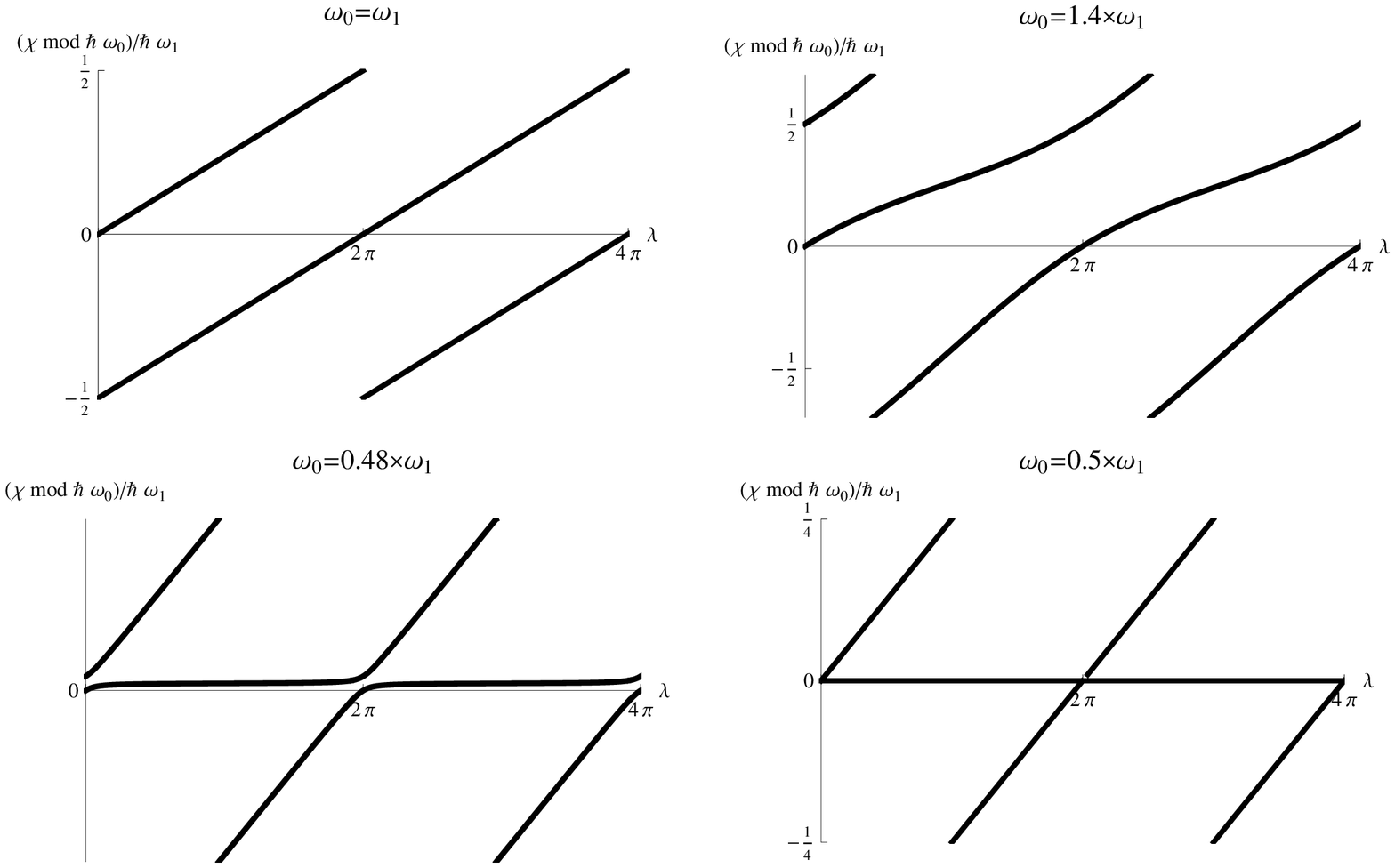}{Trajectories of the quasienergies (modulo $\hbar \omega_0$) of the Hamiltonian (\ref{model}) with respect to $\lambda \in [0,4\pi]$ for different values of $\omega_0$. The quasienergies (modulo $\hbar \omega_0$) are not $2\pi$-periodic with respect to $\lambda$ but $4 \pi$-periodic (except for $\omega_0 = \frac{\omega_1}{2}$ because of the quasienergy crossing). This is the manifestation of the Cheon's anholonomy. For $\omega_0$ in the neighbourhood of $\frac{\omega_1}{2}$ the avoided crossing restores the Cheon's anholonomy.}{14cm}
In the following, we do not consider the particular case $\omega_0 = \frac{\omega_1}{2}$, not only because the Cheon's anholonomy is absent, but also because the adiabatic approximation is not valid for this case (because of the crossing).

\subsection{The gerbe}
Let $M = S^1$ be the circle parametrized by $\lambda \mod 4 \pi$. The quasienergies are not continuous at $\lambda = 0 \mod 4 \pi$ but the discontinuity is just a Floquet block change. Let $\{U^\alpha \}_{\alpha = 1,2,3}$ be the good open cover of $S^1$ defined by the figure \ref{coverS1}.
\fig{coverS1}{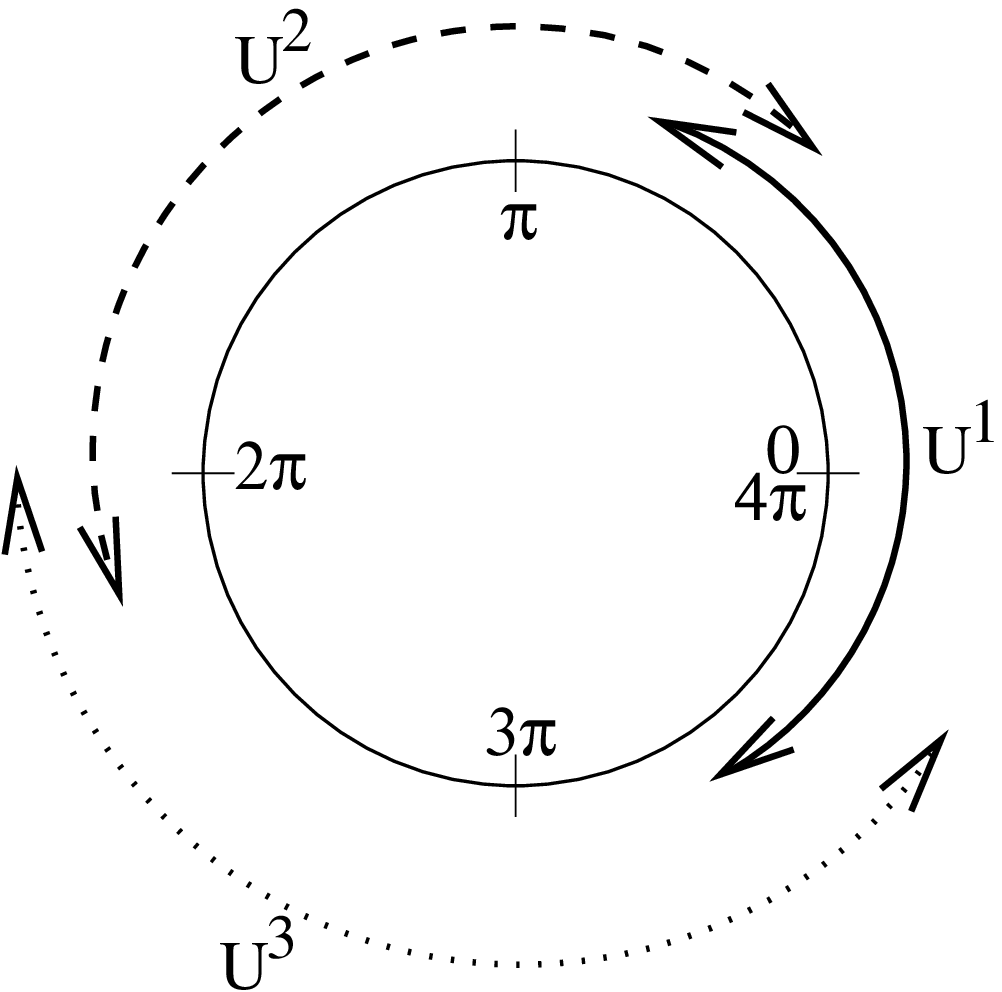}{$M = S^1$ the parameter space spanned by $\lambda \mod 4 \pi$ and its good open cover $\{U^\alpha \}_{\alpha=1,2,3}$.}{6cm}
Let $\ell^\alpha$ be a coordinate system on $U^\alpha$. $\lambda \mod 4 \pi$ being assimilated to a geometric point of $S^1$, we have
\begin{eqnarray}
\forall \lambda \in U^1, & \quad & \ell^1(\lambda) \in ]- \pi, \pi [ \\
\forall \lambda \in U^2, & \quad & \ell^2(\lambda) \in ]0 , 3 \pi [ \\
\forall \lambda \in U^3, & \quad & \ell^3(\lambda) \in ]2 \pi, 4 \pi[
\end{eqnarray}
We set 
\begin{eqnarray}
\forall \lambda \in U^\alpha, & \quad & \chi_{2n+1}^\alpha(\lambda) = \ell^\alpha(\lambda) \frac{\hbar \omega_0}{4 \pi} + n \hbar \omega_0, \quad n \in \mathbb Z \\
\forall \lambda \in U^\alpha, & \quad & \chi_{2n+2}^\alpha(\lambda) = \ell^\alpha(\lambda) \frac{\hbar \omega_0}{4 \pi} + \frac{\hbar \omega_0}{2} + n \hbar \omega_0, \quad n \in \mathbb Z
\end{eqnarray}
$\lambda \mapsto \chi^\alpha_a(\lambda)$ is then a continuous function on $U^\alpha$. Since
\begin{eqnarray}
\forall \lambda \in U^1 \cap U^2, & \quad & \ell^2(\lambda) - \ell^1(\lambda) = 0 \\
\forall \lambda \in U^2 \cap U^3, & \quad & \ell^3(\lambda) - \ell^2(\lambda) = 0 \\
\forall \lambda \in U^3 \cap U^1, & \quad & \ell^3(\lambda) - \ell^1(\lambda) = 4 \pi
\end{eqnarray}
we then have
\begin{eqnarray}
\forall \lambda \in U^1 \cap U^2, & \quad & \chi^2_a(\lambda) = \chi^1_a(\lambda) \\
\forall \lambda \in U^2 \cap U^3, & \quad & \chi^3_a(\lambda) = \chi^2_a(\lambda) \\
\forall \lambda \in U^1 \cap U^3, & \quad & \chi^3_a(\lambda) = \chi^1_a(\lambda) + \hbar \omega_0
\end{eqnarray}
This ensures that starting from ${\ell^1}^{-1}(0)$ with $\chi^1_a({\ell^1}^{-1}(0))$ and following $S^1$ we arrive after one turn with $\chi^1_a({\ell^1}^{-1}(0)) + \hbar \omega_0 = \chi^1_{a+2}({\ell^1}^{-1}(0))$. With these local definitions of the quasienergies, we respect the properties described in the previous paragraph. We conclude that
\begin{equation}
n^{12} = 0 \qquad n^{23} = 0 \qquad n^{13} = 1
\end{equation}
We note that $U^1 \cap U^2 \cap U^3 = \varnothing$ and then the cocycle relation with respect to the indices $1,2,3$ does not need to be satisfied.\\

The system associated with the Hamiltonian (\ref{model}) is then seen to be an example where one needs to introduce a Floquet block change at a chart intersection in order to define locally continuous quasienergies. 

\subsection{The geometric phase}
The extended parameter manifold is the torus $M_+ = T^2 = S^1 \times S^1$ generated by $\theta \mod 2 \pi$ and by $\lambda \mod 4 \pi$. We compute the quasienergy state associated with $\chi_1$ for $\omega_0 = \omega_1$. $\forall \theta \in [0,2\pi[$ we have
\begin{eqnarray}
& & Z_{\ell^\alpha(\lambda)}(\theta)  =  U_{\ell^\alpha(\lambda)}(\theta) e^{- \imath \mathsf M_{\ell^\alpha(\lambda)} \theta} \\
& = & e^{\imath \frac{\ell^\alpha(\lambda) \theta}{4 \pi}} \nonumber \\
& \times & \left(\begin{array}{cc} \cos^2 \frac{\ell^{\alpha}(\lambda)}{4} + e^{\imath \frac{\theta}{2}} \sin^2 \frac{\ell^{\alpha}(\lambda)}{4} & \frac{1}{2}(e^{\imath \frac{\theta}{2}}-1) \sin \frac{\ell^\alpha{\lambda}}{2} \\ \frac{1}{2} (1-e^{- \imath \frac{\theta}{2}}) \sin \frac{\ell^\alpha(\lambda)}{2} &  \cos^2 \frac{\ell^\alpha(\lambda)}{4} + e^{- \imath \frac{\theta}{2}} \sin^2 \frac{\ell^\alpha(\lambda)}{4} \end{array} \right)
\end{eqnarray}
and
\begin{equation}
Z_{\ell^\alpha(\lambda)}(2 \pi) = id_{\mathcal H}
\end{equation}
We then have $\forall \theta \in [0,2\pi[$
\begin{eqnarray}
& & |1(\theta),\lambda \rangle^\alpha =  Z_{\ell^\alpha(\lambda)}(\theta) |\mu_1,\ell^{\alpha}(\lambda) \rangle \\
& = & e^{- \imath \frac{\ell^\alpha(\lambda) \theta}{4 \pi}} \cos \frac{\ell^\alpha(\lambda)}{4} |\uparrow \rangle - e^{- \imath \frac{\ell^\alpha(\lambda) \theta}{4 \pi}} e^{- \imath \frac{\theta}{2}} \sin\frac{\ell^\alpha(\lambda)}{4} |\downarrow \rangle
\end{eqnarray}
and
\begin{equation}
 |1(2 \pi),\lambda \rangle^\alpha = \cos \frac{\ell^\alpha(\lambda)}{4} |\uparrow \rangle - \sin \frac{\ell^\alpha(\lambda)}{4} |\downarrow \rangle
\end{equation}
and then
\begin{eqnarray}
\eta^\alpha_M & = & \frac{1}{2 \pi} {^\alpha}\langle 1(\theta),\lambda|\partial_\lambda|1(\theta),\lambda \rangle^{\alpha}_{\mathcal H}d\lambda \\
& = & \frac{\imath}{8 \pi^2} \theta \left(1-\delta(\theta-2\pi)\right)d\lambda
\end{eqnarray}

\begin{eqnarray}
\eta^\alpha_0 & = & \frac{\omega_0}{2 \pi} {^\alpha}\langle 1(\theta),\lambda|\partial_\theta|1(\theta),\lambda \rangle^\alpha_{\mathcal H} \\
& = & \frac{\imath \omega_0}{8 \pi} \left( \frac{\ell^\alpha(\lambda)}{\pi} - \sin^2 \frac{\ell^\alpha(\lambda)}{4} \right)
\end{eqnarray}
We conclude that
\begin{eqnarray}
B^\alpha & = & \frac{\imath}{8 \pi^2} \theta(1-\delta(\theta-2\pi)) d\lambda \wedge d\theta \nonumber \\
& & \quad - \frac{\imath \omega_0}{8 \pi} \left(\frac{\ell^\alpha(\lambda)}{\pi} - \sin^2 \frac{\ell^\alpha(\lambda)}{4} \right) d\theta \wedge dt \nonumber \\
& & \quad - \frac{\imath \omega_0}{8 \pi} \left(\frac{1}{\pi}-\frac{1}{4} \sin \frac{\ell^\alpha(\lambda)}{2} \right) d\lambda \wedge dt \nonumber \\
& & \quad - \frac{\imath \omega_0}{16 \pi} \left(\frac{\ell^\alpha(\lambda)}{\pi} - \sin^2 \frac{\ell^\alpha(\lambda)}{4} \right) d\lambda \wedge dt
\end{eqnarray}

\begin{equation}
A^{\alpha \beta} = \frac{\imath}{2 \pi} (\delta^{\alpha 1} \delta^{\beta 3}- \delta^{\alpha 3} \delta^{\beta 1}) \omega_0 t (d\theta+ \frac{1}{2} d\lambda)
\end{equation}
where $\delta^{\alpha \beta}$ is the Kronecker symbol ($\delta^{\alpha \beta} = 1$ if $\alpha=\beta$ and $\delta^{\alpha \beta}=0$ if $\alpha \not=\beta$).

\begin{equation}
H = - \frac{\imath \omega_0}{8 \pi}\left(\frac{1}{\pi}- \sin \frac{\ell^\alpha(\lambda)}{2} \right) d\lambda \wedge d\theta \wedge dt
\end{equation}

Let $[0,T] \ni t \mapsto \lambda(t) = \frac{4 \pi t}{T} \in M=S^1$ be an example of a closed path. We chose $\lambda^{31} = \frac{7 \pi}{2}$ as the arbitrary transition point in $U^3 \cap U^1$. The geometric phase is then
\begin{eqnarray}
e^{\imath \gamma_1(\mathcal S)} & = & e^{ \frac{\imath}{8 \pi^2} \int_0^{7 \pi/2} \int_0^{2 \pi} \theta d\theta d\lambda} e^{\frac{\imath \omega_0}{8 \pi} \int_0^{7 T/8} \int_0^{2 \pi} (\frac{4t}{T} - \sin^2 \frac{\pi t}{T}) d\theta dt} \nonumber \\
& & \times e^{\imath \frac{7 \omega_0 T}{8}} \nonumber \\
& & \times e^{ \frac{\imath}{8 \pi^2} \int_{- \pi/2}^0 \int_0^{2 \pi} \theta d\theta d\lambda} e^{\frac{\imath \omega_0}{8 \pi} \int_{7T/8}^{T} \int_0^{2 \pi}(\frac{4t}{T}-4 - \sin^2 (\frac{\pi t}{T}-\pi)) d\theta dt}\\
& = & e^{\imath \pi} e^{\imath \frac{\omega_0 T}{4}} \\
& = & - e^{\imath \frac{\omega_0 T}{4}}
\end{eqnarray}

\section{Conclusion}
The geometric phase phenomenon of the adiabatic Floquet theory is an example of a higher gauge theory similar to the classical string theory. Geometric phases are horizontal lifts in a gerbe of surfaces which can be viewed as world-sheets of closed strings. The gerbe is topologically defined by a Dixmier-Douady class (degree three cohomology), which is the ``cup-product'' of a first Chern class (degree two cohomology) as for the usual adiabatic bundle by a cohomological class of degree one associated with the Floquet block changes. The Cheon's anholonomy is the origin of the degree one non-triviality (not any gauge transformation cancels $n^{\alpha \beta}$). As in the usual adiabatic phase theory, the degree two non-triviality (not any gauge transformation cancels $z^{\alpha \beta \gamma}$) is due to eigenvalue crossings (and the associated magnetic monopoles).\\
The adiabatic approximation is related to two scales of time. The adiabatic parameter variations are supposed to be slower than the quantum proper time of transition from an eigenstate to another one \cite{messiah}. The quantum system adapts to its environment, as characterized by the current adiabatic parameters, before these parameters change significantly. Consequently the wave function remains on the same instantaneous eigenvector during the evolution. This behaviour generates a usual geometric phase associated with a principal bundle with connection (a degree one Deligne cohomological class $(g^{\alpha \beta},A^\alpha)$). In the adiabatic Floquet theory we consider three scales of time: the slow adiabatic parameter variations, the fast quantum proper transitions, and the fast oscillations of the laser field wave (or the fast kick repetitions). The quantum system adapts to the adiabatic parameters before they change significantly, and it feels only the average effect of the fast oscillations. Consequently the geometric phase is associated with a gerbe with connection (a degree two Deligne cohomological class $(h^{\alpha \beta \gamma},A^{\alpha \beta},B^\alpha)$). We can conjecture that a quantum system presenting three different scales of time could always be associated with geometric phases related to a gerbe, and also that a quantum system presenting more than three different scales of time (for example a molecule interacting with two or more laser fields with incommensurable frequencies) is associated with geometric structures with a Deligne degree larger than two.

\ack
The author thanks Professor John P. Killingbeck for his help.\\
This work is supported by grants from \textit{Agence Nationale de la Recherche} (CoMoC project).

\appendix
\section{}
Let $f^* B^\alpha = \eta^\alpha_{M\mu}(\vec R(t),\theta) \frac{\partial R^\mu}{\partial t} dt \wedge d\theta + \eta^\alpha_0(\vec R(t),\theta) dt \wedge d\theta \in \Omega^2(\mathbb R \times S^1)$ be the 2-form generating the geometric phase.  Since $|a(\theta),\vec R(t) \rangle^\beta = e^{\imath \varphi^{\alpha \beta}(\vec R(t))} e^{\imath n^{\alpha \beta} \theta} |a(\theta),\vec R(t) \rangle^\alpha$ we have
\begin{equation}
\eta^\beta_{M \mu} \frac{\partial R^\mu}{\partial t} dt = \eta^\alpha_{M\mu} \frac{\partial R^\mu}{\partial t} dt+ \frac{\imath}{2 \pi} \frac{\partial \varphi^{\alpha \beta}}{\partial t} dt
\end{equation}
\begin{equation}
\eta^\beta_0 = \eta^\alpha_0 + \frac{\imath}{2 \pi} n^{\alpha \beta} \omega_0
\end{equation}

We have then
\begin{equation}
f^*B^\beta = f^* B^\alpha + \frac{\imath}{2 \pi} \frac{\partial \varphi^{\alpha \beta}}{\partial t} dt \wedge d\theta + \frac{\imath}{2 \pi} n^{\alpha \beta} \omega_0 dt \wedge d\theta
\end{equation}
Let $t^{\alpha \beta} < t^{\alpha \beta \prime}$ be two arbitrary times such that $\vec R(t^{\alpha \beta}),\vec R(t^{\alpha \beta \prime}) \in \mathcal C \cap U^\alpha \cap U^\beta$.
\begin{eqnarray}
& & \int^{t^{\alpha \beta\prime}} \int_0^{2 \pi} f^* B^\alpha + \int_{t^{\alpha \beta \prime}} \int_0^{2 \pi} f^* B^\beta \nonumber \\
& &  =  \int^{t^{\alpha \beta}} \int_0^{2 \pi} f^* B^\alpha + \int_{t^{\alpha \beta}} \int_0^{2 \pi} f^* B^\beta \nonumber \\
& & \qquad + \int_{t^{\alpha \beta}}^{t^{\alpha \beta \prime}} \int_0^{2 \pi} f^* B^\alpha + \int_{t^{\alpha \beta \prime}}^{t^{\alpha \beta}} \int_0^{2 \pi} f^* B^\beta \\
& & = \int^{t^{\alpha \beta}} \int_0^{2 \pi} f^* B^\alpha + \int_{t^{\alpha \beta}} \int_0^{2 \pi} f^* B^\beta \nonumber \\
& & \qquad + \int_{t^{\alpha \beta}}^{t^{\alpha \beta \prime}} \int_0^{2 \pi} (f^* B^\alpha - f^* B^\beta)
\end{eqnarray}
\begin{eqnarray}
& & \int_{t^{\alpha \beta}}^{t^{\alpha \beta \prime}} \int_0^{2 \pi} (f^* B^\alpha - f^* B^\beta) \\
& & = - \frac{\imath}{2 \pi} \int_{t^{\alpha \beta}}^{t^{\alpha \beta \prime}} \int_0^{2 \pi} \left( \frac{\partial \varphi^{\alpha \beta}}{\partial t} + n^{\alpha \beta} \omega_0 \right) dt d\theta \\
& & = \imath \left(\varphi^{\alpha \beta}(\vec R(t^{\alpha \beta})) - \varphi^{\alpha \beta}(\vec R(t^{\alpha \beta \prime})) \right) \nonumber \\
& & \qquad + \imath n^{\alpha \beta} \left( \omega_0 t^{\alpha \beta} - \omega_0 t^{\alpha \beta \prime} \right)
\end{eqnarray}
 We see that the quantity which is independent of the arbitrary choice of the transition point $\vec R(t^{\alpha \beta})$ is
\begin{equation}
\int^{t^{\alpha \beta}} \int_0^{2 \pi} f^* B^\alpha + \int_{t^{\alpha \beta}} \int_0^{2 \pi} f^* B^\beta + \imath \varphi^{\alpha \beta}(\vec R(t^{\alpha \beta})) + \imath n^{\alpha \beta} \omega_0 t^{\alpha \beta}
\end{equation}
We conclude that the correct definition of the geometric phase associated with a path crossing $U^\alpha \cap U^\beta$ is
\begin{equation}
e^{\int^{t^{\alpha \beta}} \int_0^{2 \pi} f^* B^\alpha} e^{\imath \varphi^{\alpha \beta}(\vec R(t^{\alpha \beta}))} e^{\imath n^{\alpha \beta} \omega_0 t^{\alpha \beta}} e^{\int_{t^{\alpha \beta}} \int_0^{2 \pi} f^* B^\beta}
\end{equation}

\section{}
Since $|a(\theta),\vec R \rangle^\beta = e^{\imath \varphi^{\alpha \beta}(\vec R)} e^{\imath n^{\alpha \beta} \theta} |a(\theta),\vec R \rangle^\alpha$ we have
\begin{equation}
\eta^\beta_M = \eta^\alpha_M + \frac{\imath}{2 \pi} d_M \varphi^{\alpha \beta}
\end{equation}
\begin{equation}
\eta^\beta_0 = \eta^\alpha_0 + \frac{\imath}{2 \pi} n^{\alpha \beta} \omega_0
\end{equation}
\begin{equation}
\chi^\beta_a = \chi^\alpha_a + n^{\alpha \beta} \hbar \omega_0 \Rightarrow d_M \chi^\beta_a = d_M \chi^\alpha_a
\end{equation}

Since
\begin{eqnarray}
B^\alpha & = & \eta^\alpha_M \wedge d\theta + F^\alpha_M - \eta^\alpha_0 d\theta \wedge dt -d_M \eta^\alpha_0 \wedge dt \nonumber \\
& & + \frac{2 \pi}{\hbar \omega_0} \eta^\alpha_M \wedge d_M \chi^\alpha_a + \frac{2 \pi}{\hbar \omega_0} \eta^\alpha_0 dt \wedge d_M \chi^\alpha_a
\end{eqnarray}
we have
\begin{eqnarray}
B^\beta - B^\alpha & = & \frac{\imath}{2 \pi} d_M \varphi^{\alpha \beta} \wedge \left(d\theta+\frac{2\pi}{\hbar \omega_0} d_M \chi^\alpha_a\right) \nonumber \\
& & - \frac{\imath}{2 \pi} n^{\alpha \beta} \omega_0 \left(d\theta+ \frac{2 \pi}{\hbar \omega_0} d_M \chi^\alpha_a \right) \wedge dt
\end{eqnarray}
We also have
\begin{equation}
A^{\alpha \beta} = \frac{\imath}{2 \pi} \left( \varphi^{\alpha \beta} + n^{\alpha \beta} \omega_0 t \right) \left(d\theta+ \frac{2 \pi}{\hbar \omega_0} d_M \chi^\alpha_a \right)
\end{equation}
so that
\begin{equation}
d_{M^+_+} A^{\alpha \beta} = \frac{\imath}{2 \pi} \left( d_M \varphi^{\alpha \beta} + n^{\alpha \beta} \omega_0 dt \right) \wedge \left( d\theta + \frac{2 \pi}{\hbar \omega_0} d_M \chi^\alpha_a \right)
\end{equation}
We then have $B^\beta - B^\alpha = d_{M_+^+} A^{\alpha \beta}$.

\begin{eqnarray}
A^{\beta \gamma} - A^{\alpha \gamma} + A^{\alpha \beta} & = & \frac{\imath}{2 \pi} ( \varphi^{\beta \gamma} - \varphi^{\alpha \gamma} + \varphi^{\alpha \beta}  \nonumber \\
& & \qquad  + \underbrace{(n^{\beta \gamma} - n^{\alpha \gamma} + n^{\alpha \beta})}_{= 0} \omega_0t ) \nonumber \\
& & \qquad \times \left(d\theta+ \frac{2 \pi}{\hbar \omega_0} d_M \chi^\alpha_a \right) \\
& = & \imath z^{\alpha \beta \gamma} \left(d\theta+ \frac{2 \pi}{\hbar \omega_0} d_M \chi^\alpha_a \right) \\
& = & - h^{\alpha \beta \gamma} d_{M^+_+} h^{\alpha \beta \gamma}
\end{eqnarray}

\begin{eqnarray}
d_{M^+_+} B^\alpha & = & F^\alpha_M \wedge d\theta - d_M \eta^\alpha_0 \wedge d\theta \wedge dt  + \frac{\partial F^\alpha_{M \mu \nu}}{\partial \theta} dR^\mu \wedge dR^\nu \wedge d\theta \nonumber \\
& & \quad + \frac{\partial^2 \eta^\alpha_0}{\partial \theta \partial R^\mu} dR^\mu \wedge d\theta \wedge dt \nonumber \\
& & \quad + \frac{2 \pi}{\hbar \omega_0} F^\alpha_M \wedge d_M \chi^\alpha_a \nonumber \\
& & \quad + \frac{2 \pi}{\hbar \omega_0} \frac{\partial \eta^\alpha_{M\mu}}{\partial \theta} \frac{\partial \chi^\alpha_a}{\partial R^\nu} dR^\mu \wedge dR^\nu \wedge d\theta \nonumber \\
& & \quad - \frac{2 \pi}{\hbar \omega_0} d_M \eta^\alpha_0 \wedge d_M \chi^\alpha_a \wedge dt \nonumber \\
& & \quad + \frac{2 \pi}{\hbar \omega_0} \frac{\partial \eta^\alpha_0}{\partial \theta} d_M \chi^\alpha_a \wedge d \theta \wedge dt
\end{eqnarray}

\begin{eqnarray}
\eta^\beta_M = \eta^\alpha_M + \frac{\imath}{2 \pi} d_M \varphi^{\alpha \beta} & \Rightarrow & d_M \eta^\beta_M = d_M \eta^\alpha_M \text{ and } \partial_\theta \eta^\beta_M = \partial_\theta \eta^\alpha_M \\ & \Rightarrow & F^\beta_M = F^\alpha_M
\end{eqnarray}

\begin{equation}
\eta^\beta_0 = \eta^\alpha_0 + \frac{\imath}{2 \pi} n^{\alpha \beta} \omega_0 \Rightarrow \frac{\partial \eta^\beta_0}{\partial R^\mu} = \frac{\partial \eta^\alpha_0}{\partial R^\mu}
 \text{ and } \partial_\theta \eta^\beta_0 = \partial_\theta \eta^\alpha_0 \end{equation}
This proves that $d_{M^+_+} B^\alpha = d_{M^+_+} B^\beta$ and then that $H = d_{M^+_+} B$ is indeed a globally defined 3-form.

\section{}
We want to solve the equation
\begin{equation}
\ihbar \omega_0 \frac{\partial U_\lambda}{\partial \theta} = H(\lambda,\theta) U_\lambda(\theta) \quad , \quad U_\lambda(0) = id_{\mathcal H}
\end{equation}
with
\begin{equation}
H(\lambda,\theta) = H_0 + \hbar \omega_0 \lambda W \sum_{n \in \mathbb Z} \delta(\theta - 2 n \pi) \quad , \quad W^2 = W
\end{equation}
We can formally write
\begin{equation}
U_\lambda(\theta) = \lim_{\epsilon \to 0} \Te^{- \imath \int_\epsilon^{\theta+\epsilon} \left( \frac{H_0}{\hbar \omega_0} + \lambda W \sum_{n \in \mathbb Z} \delta(\theta'-2n \pi) \right) d\theta'}
\end{equation}
$\mathbb T$ is the time-ordering operator and then $\Te^{\int}$ symbolizes the Dyson expansion. For $\theta< 2 \pi$ we have
\begin{equation}
U_\lambda(\theta) = \Te^{- \imath \int_0^\theta \frac{H_0}{\hbar \omega_0} d\theta'} = e^{- \frac{\imath}{\hbar \omega_0} H_0 \theta}
\end{equation}
For $\theta = 2\pi$, by using the intermediate representation theorem we have
\begin{eqnarray}
U_\lambda(2 \pi) & = & \lim_{\epsilon \to 0} \Te^{- \imath \int_0^{2 \pi + \epsilon} (\frac{H_0}{\hbar \omega_0} + \lambda W \delta(\theta-2 \pi)) d\theta} \\
& = & \lim_{\epsilon \to 0} \Te^{- \imath \int_0^{2 \pi} \frac{H_0}{\hbar \omega_0} d\theta} \nonumber \\
& & \qquad \times  \Te^{- \imath \int_0^{2 \pi + \epsilon} \Te^{ \imath \int_0^{\theta} \frac{H_0}{\hbar \omega_0} d\theta'} \lambda W \Te^{- \imath \int_0^{\theta} \frac{H_0}{\hbar \omega_0} d\theta'} \delta(\theta-2 \pi)d\theta} \\
& = & \lim_{\epsilon \to 0} e^{- \frac{\imath}{\hbar \omega_0} H_0 2 \pi} \Te^{\imath \int_0^{2 \pi + \epsilon} e^{\frac{\imath}{\hbar \omega_0} H_0 \theta} \lambda W e^{- \frac{\imath}{\hbar \omega_0} H_0 \theta} \delta(\theta-2 \pi) d\theta}
\end{eqnarray}

For any operator $K(\theta)$ we have
\begin{eqnarray}
& & \Te^{- \imath \int_0^{2 \pi + \epsilon} K(\theta)\delta(\theta-2 \pi)d\theta} \nonumber \\
& & =  id_{\mathcal H} + \sum_{n=1}^{+ \infty} (-\imath)^n \int_0^{2 \pi + \epsilon} K(\theta_1) \delta(\theta_1-2\pi) \int_0^{\theta_1} K(\theta_2) \delta(\theta_2-2 \pi) ...\nonumber \\
& & \quad ...  \int_0^{\theta_{n-1}} K(\theta_n) \delta(\theta_n-2 \pi) d\theta_n ... d\theta_1
\end{eqnarray}

We have for the integrals the results
\begin{equation}
\int_0^{2 \pi + \epsilon} K(\theta_1) \delta(\theta_1-2 \pi) d\theta_1 = K(2 \pi)
\end{equation}
\begin{eqnarray}
& & \int_0^{2 \pi + \epsilon} K(\theta_1) \delta(\theta_1-2 \pi) \int_0^{\theta_1} K(\theta_2) \delta(\theta_2-2\pi) d\theta_2 d\theta_1 \nonumber \\
& & = K(2 \pi)^2 \int_{\theta_1=0}^{2 \pi+\epsilon} \int_{\theta_2=0}^{\theta_1} \delta(\theta_1-2\pi) \delta(\theta_2-2\pi) d\theta_2 d\theta_1 
\end{eqnarray}
This last equation should be treated with some caution, since the product of two singular distributions is not well defined. Since the double integration refers to the domain of $[0, 2 \pi+\epsilon]^2$ defined by $0 \leq \theta_2 \leq \theta_1 \leq 2 \pi + \epsilon$ we have
\begin{eqnarray}
& & \int_{\theta_1=0}^{2 \pi + \epsilon} \int_{\theta_2=0}^{\theta_1} \delta(\theta_1-2\pi) \delta(\theta_2-2 \pi) d\theta_2 d\theta_1 \nonumber \\
& &  =  \int_{\theta_2=0}^{2 \pi+\epsilon} \int_{\theta_1=\theta_2}^{2 \pi+\epsilon} \delta(\theta_1-2 \pi) \delta(\theta_2-2 \pi) d\theta_2 d\theta_1 \\
& &  =  \int_{\theta_2=0}^{2 \pi+\epsilon} \int_{\theta_1=\theta_2}^{2\pi+\epsilon} \delta(\theta_2-2 \pi) \delta(\theta_1-2\pi) d\theta_2 d\theta_1 \\
& &  =  \int_{\theta_1=0}^{2 \pi+\epsilon} \int_{\theta_2=\theta_1}^{2\pi+\epsilon} \delta(\theta_1-2\pi) \delta(\theta_2-2\pi) d\theta_1 d\theta_2
\end{eqnarray}
We then have
\begin{eqnarray}
& & \int_{\theta_1=0}^{2\pi+\epsilon} \int_{\theta_2=0}^{\theta_1} \delta(\theta_1-2\pi) \delta(\theta_2-2\pi) d\theta_2 d\theta_1 \nonumber \\
& = &  \frac{1}{2} \left( \int_{\theta_1=0}^{2\pi+\epsilon} \int_{\theta_2=0}^{\theta_1} \delta(\theta_1-2\pi) \delta(\theta_2-2\pi) d\theta_2 d\theta_1 \right. \nonumber \\
& & \quad + \left. \int_{\theta_1=0}^{2\pi+\epsilon} \int_{\theta_2=\theta_1}^{2\pi+\epsilon} \delta(\theta_1-2\pi) \delta(\theta_2-2\pi) d\theta_1 d\theta_2 \right) \\
& = & \frac{1}{2} \int_0^{2\pi+\epsilon} \int_0^{2\pi+\epsilon} \delta(\theta_1-2\pi) \delta(\theta_2-2\pi) d\theta_1 d\theta_2 \\
& = & \frac{1}{2} \left( \int_0^{2\pi+\epsilon} \delta(\theta-2\pi) d\theta \right)^2
\end{eqnarray}
This last integration is well defined : $\int_0^{2 \pi+\epsilon} \delta(\theta-2\pi) d\theta = 1$. We conclude that
\begin{equation}
\int_0^{2 \pi+\epsilon} K(\theta_1) \delta(\theta_1-2\pi) \int_0^{\theta_1} K(\theta_2-2\pi) \delta(\theta_2) d\theta_2 d\theta_1 = \frac{K(2\pi)^2}{2}
\end{equation}
By similar demonstrations we have $\forall n \in \mathbb N^*$
\begin{equation}
\int_0^{2\pi+\epsilon} K(\theta_1) \delta(\theta_1-2\pi) \int_0^{\theta_1}...\int_0^{\theta_{n-1}} K(\theta_n) \delta(\theta_n-2\pi) d\theta_n...d\theta_1 = \frac{K(2\pi)^n}{n!}
\end{equation}
We conclude that
\begin{eqnarray}
\Te^{- \imath \int_0^{2\pi+\epsilon} K(\theta)\delta(\theta-2\pi)d\theta}  & = & id_{\mathcal H} + \sum_{n=1}^{+ \infty} \frac{(- \imath K(2\pi))^n}{n!} \\ & = & e^{- \imath K(2\pi)}
\end{eqnarray}
We can now return to the original problem:
\begin{eqnarray}
& & \lim_{\epsilon\to 0}\Te^{- \imath \int_0^{2\pi+\epsilon} e^{ \imath \frac{H_0}{\hbar \omega_0} \theta} \lambda W e^{- \imath \frac{H_0}{\hbar \omega_0} \theta} \delta(\theta-2\pi)d\theta} \nonumber \\
& & \quad   =  e^{- \imath e^{\frac{\imath}{\hbar \omega_0} H_0 2\pi} \lambda W e^{\frac{-\imath}{\hbar \omega_0} H_0 2\pi}}
\end{eqnarray}
and then
\begin{eqnarray}
U_\lambda(2\pi) & = & e^{- \frac{\imath}{\hbar \omega_0} H_0 2\pi} e^{- \imath e^{\frac{\imath}{\hbar \omega_0} H_0 2\pi} \lambda W e^{\frac{-\imath}{\hbar \omega_0} H_0 2\pi}} \\
& = & e^{- \imath \lambda W} e^{- \frac{\imath}{\hbar \omega_0} H_0 2\pi}
\end{eqnarray}
Moreover we have
\begin{eqnarray}
e^{- \imath \lambda W} & = & id_{\mathcal H} + \sum_{n=1}^{\infty} \frac{(- \imath \lambda)^n}{n!} W^n \\
& = & id_{\mathcal H} + \sum_{n=1}^{\infty} \frac{(- \imath \lambda)^n}{n!} W \\
& = & id_{\mathcal H} +\left(e^{- \imath \lambda}-1 \right) W
\end{eqnarray}
so that, finally, we have
\begin{equation}
U_\lambda(\theta) = \left\{ \begin{array}{ll} e^{- \frac{\imath}{\hbar \omega_0} H_0 \theta} & \text{if } \theta \in [0,2\pi[ \\
 \left(id_{\mathcal H} +\left(e^{- \imath \lambda}-1 \right) W \right) e^{- \frac{\imath}{\hbar \omega_0} H_0 2\pi} & \text{if } \theta = 2 \pi
\end{array} \right.
\end{equation}

\end{document}